\documentclass[prd,
               twocolumn,
               eqsecnum,
               showpacs,
               letterpaper,
               superscriptaddress,
               altaffilletter]{revtex4-1}
 \usepackage[utf8x]{inputenc}
 \usepackage{graphicx}
  \usepackage{amsmath,amssymb} 
 \usepackage[english]{babel}
  \usepackage{rotating}
  \usepackage{amsbsy}
 \usepackage{textcomp}
 \usepackage{psfrag}
 \usepackage{multirow}
 \usepackage{color} 
 \usepackage{sidecap}
 \usepackage{array}
 \newcommand{\ud}{\mathrm{d}}    %%%%%%%%%%%%%%%%%%%%%%%%%%%%%%%%%%%%%%%%% NUOVI COMANDI
\begin{document}

\title{Parameter estimation from gravitational waves generated by non-spinning binary black holes
with laser interferometers: beyond the Fisher information.}
\author{S. Vitale}
\affiliation{Embry-Riddle Aeronautical University, 3700 Willow Creek Road, Prescott, AZ, 86301,USA}
\affiliation{LPTMC, Universit\'e Pierre-et-Marie-Curie - 4, Place Jussieu, 75005 Paris, France}
\affiliation{Nikhef, Science Park 105, 1098 XG Amsterdam, The Netherlands}
\author{M. Zanolin}
\affiliation{Embry-Riddle Aeronautical University, 3700 Willow Creek Road, Prescott, AZ, 86301,USA}
   	   
\begin{abstract}
  In this paper we apply to gravitational waves from non-spinning binary systems a recently introduced frequentist methodology to calculate analytically
  the error for a maximum likelihood estimate (MLE) of physical parameters. While existing literature focuses on using the Cramer Rao Lower bound (CRLB) and Monte Carlo simulations, we use a power expansion of the bias and covariance in inverse powers of the signal to noise ratio. The use of higher order derivatives of the likelihood function 
  in the expansions makes the prediction also sensitive to the secondary lobes of the MLE probability distribution. 
  We discuss conditions for validity of the CRLB and predict new features in regions of the parameter space currently not explored. For example, we see how the bias can become the most important 
  contributor to the parameters' errors for high mass systems ($200 M_\odot$ and above). 
\end{abstract}

\maketitle

\newpage

\section{Introduction}

Coalescing black holes binaries (BBH) are among the most promising sources of
gravitational waves (GW) transients \cite{RATE}.
Observations strongly indicate the existence of stellar-mass 
BHs (3-10 $M_\odot$ ) \cite{bhev1}  and super massive black holes ($10^4-10^{10} M_\odot $)
(\cite{bhev2},\cite{bhev3}), also suggesting the possibility of intermediate BHs (\cite{bhev4}).
Observations and models also point to the formation of binary BH systems (BBH)
(\cite{aji5},\cite{aji6},\cite{aji7},\cite{aji8},\cite{aji9},\cite{aji10},\cite{aji11},\cite{aji12},\cite{aji13}).   
The chances of observation of BBH GWs in the next few years are promising given the 
advanced generation of laser interferometers currently 
under construction (\cite{adl},\cite{adl2}) and  recent advances in modeling the BBH waveforms. 
In fact, even if the inspiral and ringdown phases of the life of a binary system 
are well understood (the inspiral GW can be computed using Post-Newtonian approximations 
(\cite{IMR-Blanchet2006}), while the GW for the ringdown phase can be obtained with black hole perturbation theories)
only recently numerical breakthroughs in numerical relativity made compute GWs from the merger phase possible (\cite{IMR-Pretorius2005}, \cite{IMR-Campanelli2006}, \cite{IMR-Baker2006}, \cite{IMR-Herrmann2007}, \cite{IMR-Sperhake2006}, \cite{IMR-Brugmann2006}, \cite{IMR-Thornburg2007}, \cite{IMR-Etienne2007}).
The GW community is actively preparing for detection and parameter estimation 
opportunities in GWs from BBHs \cite{ninja} and plans are shaping up to explore all the 
interesting regions of the parameter space. P. Ajith and collaborators (\cite{IMR-Ajith2007}) 
have proposed a template bank for the waveform coming from a coalescing binary system, 
made of non-spinning masses, that takes into account the inspiral merger and ringdown stages 
of the binary's life (IMR).
IMR waveforms are obtained by tuning numerical parameters of a phenomenological wave (see \cite{IMR-Ajith2007})  to a set of numerical calculations performed in the full GR. Other attempts to produce analytical approximations have been developed in more recent times, like the effective one body model (EOB) proposed in \cite{Buonanno2007}, that was used in the fifth LIGO run S5. The most accurate EOB model in the literature is described in (\cite{Damour2009}, \cite{Buonanno2009}). The discrepancies between different analytical approximations to numerical relativity waveforms are described in \cite{Santamaria2010} and show for example that the amplitude of the IMR waveforms can be up to 20\% different from the EOB ones. This means that a binary at a given distance could produce a signal to noise ratio (SNR) at receiver up to 20\% different from the one expected using IMR waveforms. 

This paper discusses the estimation accuracy of BBH physical parameters which can be obtained with advanced configuration laser interferometers and IMR waveforms. Existing literature \cite{IMR-AjithApril2009}  predicts the errors with approaches that have some intrinsic limitations:
(a) with the square root of the inverse of the Fisher information matrix elements (commonly named
Cramer Rao Lower Bound (CRLB), \cite{IMR-Helstrom1968}) that only takes into account the curvature 
of the likelihood function around the true value of the parameters.
The CRLB is known to underestimate the error in low signal to noise ratio (SNR)
 (\cite{IMR-Vecchio1998}, \cite{IMR-Bala1996}, \cite{IMR-Bala1998})
and is the lowest possible uncertainty for unbiased estimators, or (b)
perform simplified Monte Carlo simulations. In the MC simulations  
\cite{IMR-AjithApril2009} $\eta$ is enforced to values $<= 0.25$ and the
simulations do not explore the secondary peaks of the likelihood  
function.

A recent paper, \cite{IMR-ZanolinVitale2010}, from the authors, studies 
the problem of MLE errors from GWs from the inspiral phase of binary systems 
with asymptotic expansions for the covariance and the bias in terms of  
 power series in the inverse of the SNR.    
The first order of the covariance series is the inverse of the Fisher information matrix. The second order is a more complicated expression that depends on the secondary maximum of the 
probability distribution because it contains higher order derivatives of the likelihood function (up to the fourth).
The first two orders of the covariance and bias expansion are 
a better tool than the CRLB to estimate the errors and allow the determination of necessary conditions for the validity of the CRLB (for example
by requiring the second order to be much smaller than the first).
These conditions allow predictions to be made on the interferometers' capability to estimate parameters in different regions of the parameter space. 
We show that the variation of the errors in the parameter space 
is more complicated than as predicted solely by the CRLB and that 
the existence of minima for particular values of the masses 
can happen even without including the merger phase.
We also predict that the bias  can become the most important contributor 
to the parameters' errors for high mass systems ($200 M_\odot$ and above), 
due to the nonlinear dependence of the signal on the parameters
(these regions of the parameter space are not yet explored).

In section \ref{Section_DataAnalysis} we define the data model, the MLE, the statistical errors and how to compute the asymptotic expansions.
In section \ref{Section_IMR} we give the analytical expression for the IMR GW for BBHs,
and the noise spectra for the advanced configurations of LIGO and Virgo.
In section \ref{Section_Results} the results are described. 
\section{Statistical model}\label{Section_DataAnalysis}
We model the output $x(t)$ of a GW detector as the sum of the GW signal $h(t,\theta^\mu)$, that depend on the vector of unknown parameters $\theta^\mu$, and 
a Gaussian stationary noise $w(t)$ with zero mean, $E[w(t)]=0$, with $E[A]$ denoting the mean of $A$ over the ensemble  
%\label{signal_time}
\begin{equation}\label{signal_time}
x(t) = h(t,\theta^\mu) + w(t)
\end{equation}
%\label{noise_mean}
%\begin{equation}\label{noise_mean}
% E[w(t)]=0 
%\end{equation}
If one expects the wave to have a known analytical form it is possible to perform an estimation of the parameters by filtering detector data with a bank of waveform templates. In Gaussian noise, this is a maximum
likelihood estimation.\\
Given the Fourier transform  for a function $h(t)$ as  $h(f)\equiv \int{dt e^{- 2 \pi i f t} h(t)}$,
the expectation of the product of two functions can be written like a scalar product
\begin{equation}\label{IMR-MeanFreq}
\langle u(f)\,|\,v(f) \rangle \equiv 2 \int_{f_{low}}^{f_{cut}}{ \ud f\, \frac{u(f) v(f)^{*} + u(f)^{*} v(f) }{S_h(f)}}
\end{equation}
where the range of integration depends on the antenna properties and on the theoretical model for the binary system, and where we introduced the {\it one sided noise spectral density}, $S_h(f)$ defined by $E[w(f) w(f')] = \frac{1}{2} S_h(f) \delta(f-f')$.
The SNR corresponding to the optimal filter is defined as
\begin{equation}\label{optimal_snr}
\rho^2 \equiv \langle h(f)\,,\,h(f)\rangle = 4 \int_{f_{low}}^{f_{cut}}{\ud\!f \frac{|h(f)|^2}{S_h(f)}}
\end{equation}
%\subsection{Parameter estimation, CRLB and beyond}\label{Section_ParameterEstimation}
Once the values of the parameters are estimated using matched filters,
 the accuracy can be evaluated  with the square root of the 
mean squared error (MSE) for the j-th parameter:
\begin{eqnarray}
 MSE_{\vartheta^j}&\equiv& E\left[\left(\hat{\vartheta^j}- E(\hat{\vartheta^j})\right)^2\right] + \left(E\left[\hat{\vartheta^j} - \vartheta^j\right]\right)^2 \nonumber\\
& \equiv& \sigma_{\vartheta^j}^2 + b_{\vartheta^j}^2 \label{IMR-MSE}
\end{eqnarray}
For large SNRs one can use the CRLB to obtain a lower bound for the error of the j-th parameter $ MSE_{\vartheta^j} \geq \left[i^{-1}\right]_{j\,j}$,
where $i$ is the {\it Fisher Information} matrix, whose  $(j\,k)$ element can be written as a scalar product of signal's first derivatives $i_{j\,k} \equiv \langle h(f)_j\,,\,h(f)_k \rangle$ where $h(f)_j \equiv \frac{\partial h(f)}{\partial \vartheta^j}$. The above notation allows for more concise formulae:
$ \langle h_{a\,b\cdots m}\,,\,h_{n\,o\cdots z} \rangle \equiv \langle a\,b\cdots m|n\,o\cdots p \rangle $, where the f dependence is not explicitly shown.
The scalar product is the one defined in eq. \ref{IMR-MeanFreq}.
In \cite{IMR-ZanolinVitale2010} expansions for both the covariance and the bias 
like power series on $1/\rho$ are given as:
%\begin{widetext}
\begin{eqnarray}
\sigma_{\vartheta^i}^2 &=& \frac{S_1^2}{\rho^2} +\frac{S_2^2}{\rho^4}+\cdots= \sigma^2_{\vartheta^i}[1]+\sigma^2_{\vartheta^i}[2] +\cdots \label{IMR-Sigma_series}\\ \nonumber \\
b_{\vartheta^i} &=& \frac{B_1}{\rho} + \frac{B_2}{\rho^2} + \cdots=b_{\vartheta^i}[1] +b_{\vartheta^i}[2] + \cdots \label{IMR-bias_series}
\end{eqnarray}
%\end{widetext}
where the first order covariance is the CRLB, and the second order can be written as:
\begin{widetext}
\begin{eqnarray}\label{IMR-VarMatrixSimplified}	
\sigma^2_{\vartheta^j}[2] &=& i^{jm}i^{jn}i^{pq}(\upsilon_{nmpq}\!+\!3 \langle nq\,|\,pm\rangle + 2\upsilon_{nmp,q}\!+\!
\upsilon_{mpq,n}) +\\
&+& i^{jm}i^{jn}i^{pz}i^{qt}\bigg( v_{npm} v_{qzt} + \frac{5}{2} v_{npq} v_{mzt} + 2 v_{qz\,,\,n}v_{mtp} +2 v_{qp,z} v_{nmt} +\nonumber \\
&+&  6 v_{mqp}v_{nt\,,\,z} + v_{pqz} v_{nt\,,\,m} + 2 v_{mq\,,\,z} v_{pt\,,\,n} +2 v_{pt\,,\,z}v_{mq\,,\,n} + v_{mz\,,\,t}v_{nq\,,\,p}\bigg)\nonumber 
\end{eqnarray}
\end{widetext}
where $${\upsilon_{a_1\cdots a_s ,b_1 \cdots b_p,\cdots , z_1\cdots z_q}=E\left[ h_{a_1 \cdots a_s}\,h_{b_1\cdots b_p}\,\cdots h_{z_1\cdots z_q} \right] }$$ are expectations of product of the signal derivatives in the time domain. For example: $\upsilon_{a\,b,c,d}= E[h_{a\,b}\,h_c\,h_d]$.
Using some algebra, the following explicit expressions can be given, in the frequency space:

\begin{eqnarray}
\upsilon_{a,b}&=& - \upsilon_{ab} = i_{ab}=\langle a\,|\,b \rangle \label{IMR-vab} \\
\upsilon_{ab\,,\,c} & =& \langle ab\,|\,c\rangle\\
\upsilon_{abc\,,\,d}  &=& \langle abc\,|\,d\rangle\\
% \end{eqnarray}
% \begin{eqnarray}
\upsilon_{abc} &=& -\langle ab\,|\,c \rangle - \langle ac\,|\,b \rangle-\langle bc\,|\,a \rangle\\
\upsilon_{ab\,,\,cd}&=& \langle ab\,|\,cd \rangle  + \langle a\,|\,b \rangle \langle c\,|\,d \rangle\\
% \end{eqnarray}
% \begin{eqnarray}
\upsilon_{abcd}&=& -\langle ab\,|\,cd \rangle - \langle ac\,|\,bd \rangle -\langle ad\,|\,bc \rangle -\nonumber \\
&&-\langle abc\,|\,d \rangle-\langle abd\,|\,c \rangle-\langle acd\,|\,b \rangle-\langle bcd\,|\,a \rangle\nonumber \\
\upsilon_{ab\,,\,c\,,\,d}&=& -\langle a\,|\,b \rangle \,\langle c\,|\,d \rangle =-i_{ab}\,i_{cd} \label{IMR-vabcd}\\
% \end{eqnarray}
% \begin{eqnarray}
\upsilon_{abc\,,\,de} &=& \langle abc\,|\,de\rangle - i_{de} v_{abc}\\
\upsilon_{abcd\,,\,e} &=& \langle abcd\,|\,e\rangle \\
\upsilon_{abc\,,\,d\,,\,e} &=&  i_{de} v_{abc}
\end{eqnarray}
\begin{eqnarray}
&&\upsilon_{ab\,,\,cd\,,\,e}= - i_{ab} v_{cd\,,\,e} - i_{cd} v_{ab\,,\,e}\\
&&\upsilon_{abcde} = -\langle abcd\,|\,e\rangle -\langle abce\,|\,d\rangle-\langle abde\,|\,c\rangle
-\langle acde\,|\,b\rangle \nonumber \\
&&- \langle bcde\,|\,a\rangle -\langle abc\,|\,de\rangle -\langle abd\,|\,ce\rangle
-\langle acd\,|\,be\rangle \nonumber \\
&&- \langle bcd\,|\,ae\rangle -\langle abe\,|\,cd\rangle-\langle ace\,|\,bd\rangle
-\langle bce\,|\,ad\rangle \nonumber \\
&&-\langle ade\,|\,bc\rangle -\langle bce\,|\,ac\rangle -\langle cde\,|\,ba\rangle \label{IMR-vabcde}
\end{eqnarray}
Where $i_{jk}$ is the Fisher information matrix.

The reason for the presence of $3 \langle nq\,|\,pm\rangle$ in (\ref{IMR-VarMatrixSimplified}) is that this expression is a simplified version of the eq. 2.4 we gave in \cite{IMR-ZanolinVitale2010}, in which some of the $\upsilon$ have been replaced with their values in terms of scalar product. This allows for slightly shorter expression. The same kind of simplifications can be performed on the first and second orders of the bias, that have a final form shorter than that presented in \cite{IMR-ZanolinVitale2010}:
\begin{eqnarray}\label{IMR-BiasOne}
b_{\vartheta^r}[1]&=&\frac{1}{2}i^{ra}i^{bc}(\upsilon_{abc}+2\upsilon_{c,ab})
\end{eqnarray}
\begin{widetext}
\begin{eqnarray}\label{IMR-BiasTwoSimplified}
b_{\vartheta^m}[2] &=& \frac{i^{ma}i^{bd}i^{ce}}{8}[v_{abcde} + 4 \langle ac\,|\,bde\rangle + 8 \langle de \,|\,abc\rangle + 4v_{abce,d}] \nonumber\\
&+&\frac{i^{ma}i^{bc}i^{df}i^{eg}}{4}\bigg[(2v_{afed}v_{gb,c} + 2v_{bedf}v_{ac,g} + 4v_{abed}v_{gf,c}) + (v_{afed}v_{gcb} +\nonumber\\
&+& 2v_{abed}v_{gcf} + 2v_{dbeg}v_{acf}) + ( 2 v_{aed} \langle gb\,|\,fc\rangle + 4 v_{acf} \langle dg\,|\,eb\rangle + 4 v_{bed} \langle ac \,|\,gf\rangle\nonumber  \\
&+&2v_{fcb} \langle ag\,|\,ed\rangle ) + (4v_{afe,g}v_{db,c} + 4v_{afe,c}v_{db,g} + 4v_{dbe,g}v_{af,c}) + (2v_{abe,g}v_{cdf}\nonumber\\
&+& 4v_{dbe,g}v_{acf} + 4v_{abe,f} v_{cdg} + 2v_{dge,b} v_{acf}) + (4 \langle ag\,|\,fc\rangle \, v_{ed,b} + 4 \langle ed\,|\,fc\rangle \, v_{ag,b} \nonumber \\
&+& 4 \langle ag\,|\,ed\rangle \,v_{fc,b}) \bigg] \nonumber \\
&+& \frac{i^{ma}i^{bc}i^{de}i^{fg}i^{ti}}{8}[v_{adf}(v_{ebc}v_{gti} + 2v_{etc}v_{gbi} + 4v_{gbe}v_{tci} + 8v_{gbt}v_{eci} + 2v_{ebc}v_{gt,i} \nonumber \\
&+& 4v_{etc}v_{gb,i} + 2v_{gti}v_{eb,c} + 4v_{gtc}v_{eb,i} + 8v_{gbt}v_{ce,i} + 8v_{gbt}v_{ci,e} + 8v_{gbe}v_{ct,i} + 8v_{cte}v_{gb,i} \nonumber \\
&+& 4v_{cti}v_{gb,e} + 4v_{gt,i}v_{eb,c} + 4v_{eb,i}v_{gt,c} + 8v_{gt,b}v_{ic,e} + 8v_{gt,e}v_{ic,b} + 4v_{bet}v_{g,c,i}) \nonumber \\
&+& v_{dci}(8v_{bgt}v_{ae,f} + 4v_{bgf}v_{ae,t} + 8v_{ae,t}v_{bg,f} + 8v_{ae,f}v_{bg,t} + 8v_{af,b}v_{ge,t})]
\end{eqnarray} 
\end{widetext}

\section{The IMR waveform}\label{Section_IMR}

IMR waveforms were obtained by tuning numerical parameters of a phenomenological wave to a set of numerical calculations performed in the full GR, \cite{IMR-Ajith2007}. Afterwards, the faithfulness of the IMR waves has been improved (\cite{IMR-Ajith2008}, \cite{IMR-AjithApril2009}, \cite{IMR-AjithMay2009}) and they have been used for the purpose of parameter estimation. In this work we follow the model presented in \cite{IMR-AjithApril2009}. The IMR waveform is written directly in the Fourier space (while the EOB are calculated in the time domain), as a piecewise function with a part for the inspiral, one for the merger, and another for the ringdown phase. Explicitely, we write the GW in the Fourier space as: 

\begin{equation}\label{Wave_Fourier}
h(f) = A_{eff}(f)\, e^{i \Psi_{eff}(f)} 
\end{equation}

where the phase and the amplitude are expressed as:
\begin{widetext}
\begin{eqnarray}
A_{eff}(f) &\equiv& \mathcal{A} \;f_{merg}^{-7/6} \left\{ 
\begin{matrix}
 \left(f/f_{merg}\right)^{-7/6} &\mbox{if}& &f&<f_{merg}\\
 \left(f/f_{merg}\right)^{-2/3} &\mbox{if}& f_{merg}\leq& f&<f_{ring}\\
 \omega \mathcal{L}(f,f_{ring},\sigma) &\mbox{if}& f_{ring}\leq &f&<f_{cut}
\end{matrix}\right. \\ \nonumber \\
\Psi_{eff}(f) &\equiv&  2 \pi f t_a + \phi_a + \frac{1}{\eta} \sum_{k\, in\, \{0,2,3,4,6\}} \left(x_k \eta^2 +y_k \eta +z_k\right) \left(\pi M f\right)^{\frac{k-5}{3}} \;.
\end{eqnarray}
\end{widetext}
We have defined:
\begin{equation}
 \omega \equiv \frac{\pi\sigma}{2} \left( \frac{f_{ring}}{f_{merg}}\right)^{-\frac{2}{3}}
\end{equation}
\begin{equation}
 \mathcal{L}(f,f_{ring},\sigma) \equiv \frac{1}{2\pi} \frac{\sigma}{(f-f_{ring})^2 +\sigma^2/4}
\end{equation}
The phenomenological parameters $f_{merg},f_{ring},\sigma,f_{cut}$  depends on the total and symmetrized mass, via the following expressions:
\begin{equation*}
f_{i}\equiv \frac{a_i \eta^2 + b_i \eta + c_i}{\pi M}
\end{equation*}
% \begin{eqnarray*}
%  f_{merg}&\equiv& \frac{a_0 \eta^2 + b_0 \eta + c_0}{\pi M} \nonumber \\  f_{ring} &\equiv & \frac{a_1 \eta^2 + b_1 \eta + c_1}{\pi M}\\
% \sigma&\equiv& \frac{a_2 \eta^2 + b_2 \eta + c_2}{\pi M}   \nonumber \\ f_{cut}&\equiv& \frac{a_3 \eta^2 + b_3 \eta + c_3}{\pi M}
% \end{eqnarray*}
with  $i=[merg,ring,\sigma,cut]$ (for the values of the numerical coefficients $a,b,c,x,y,z$ see \cite{IMR-Ajith2008}).
They represent the frequency at which the system passes from its inspiral phase to the merger ($f_{merg}$), from the merger to the ringdown ($f_{ring}$), and the frequency for which the signal ceases to be described by this model ($f_{cut}$, this is also the upper limit of the integrals (\ref{IMR-MeanFreq})).
This signal depends on five physical parameters $(\mathcal{A}, t_a,\phi_a, M,\eta)$: (I) $\mathcal{A}$ is the amplitude of the wave. It can be expressed as $\mathcal{A}= \frac{M^{5/6}}{d \, \pi^{2/3}}\sqrt{\frac{5 \eta}{24}}$, where $d$ is the effective distance of the binary.
(II) $t_a$ is the arrival time of the GW at the detector.
(III) $\phi_a$ is the arrival phase, i.e. the phase of the signal at the time $t_a$.
(IV) $M$ is the total mass of the binary.
(V) $\eta$ is the symmetrized mass ratio: $\eta\equiv m_1\,m_2/M^2$.
Sometime the chirp mass is used in the literature, instead of the total mass. They are related by $M_{chirp} = \eta^\frac{3}{5} M$.
\\
If one considers the merger and ringdown phase too, the amplitude is not uncoupled from the other parameters. One can not work in the simplified four-dimensional parameter space obtained by treating the amplitude as a known constant (\cite{IMR-ZanolinVitale2010}, \cite{IMR-Cutler1994},\cite{IMR-Arun2005}), and the full five dimensional space must be considered.

%\section{Noises}\label{Section_Noises}
We perform the calculations using either the design advanced Ligo (AdvLIGO) or advanced Virgo (AdvVirgo) noises, as they are given in {\cite{IMR-AjithApril2009}. 
The AdvLigo one sided noise spectral density is written as:
\begin{eqnarray}\label{IMR-AdvLigo}
 S_h(f)&=& S_0 \left[x^{-4.14} -5 x^{-2} + 111\frac{1-x^2+x^4/2}{1+x^2/2}\right],\; f\geq f_{low}\nonumber\\
 S_h(f)&=&\infty, \;\;f\leq f_{low}
 \end{eqnarray}
 where the lower frequency cutoff value is $f_{low}=10\mbox{Hz}$, $x\equiv \frac{f}{f_0}$, $f_0= 215 \mbox{Hz}$, and $S_0= 10^{-49} \mbox{Hz}^{-1}$.
While for the AdvVirgo:
\begin{eqnarray}\label{AdvVirgo}
 S_h(f)&=& S_0 \bigg	[ 2.67\,10^{-7}\, x^{-5.6} +0.68\, e^{-0.73\, (\ln{x})^2 } x^{5.34} \nonumber\\
 &+& 0.59\, e^{(\ln{x})^2\,\left[-3.2 -1.08 \ln{x} -0.13 (\ln{x})^2\right]} x^{-4.1} +\nonumber \\
 &+& 0.68\, e^{-0.73\, (\ln{x})^2 } x^{5.34}\bigg],\; f\geq f_{low}\nonumber\\
 S_h(f)&=&\infty, \;\;f\leq f_{low}
 \end{eqnarray}
 where the lower frequency cutoff value is chosen to be $f_{low}=10\mbox{Hz}$, $x\equiv \frac{f}{f_0}$, $f_0= 720 \mbox{Hz}$, and $S_0= 10^{-47} \mbox{Hz}^{-1}$.
Fig. \ref{Noises-IMR} shows the value of $\sqrt{S_h(f)}$ for both detectors.

\begin{figure}[htb]
\centering
 \begin{tabular}{lc}
\multirow{1}{10mm}[20mm]{\small ${S_h(f)}^{\frac{1}{2}}$} & \includegraphics[width=5.5 cm,height=4.5cm]{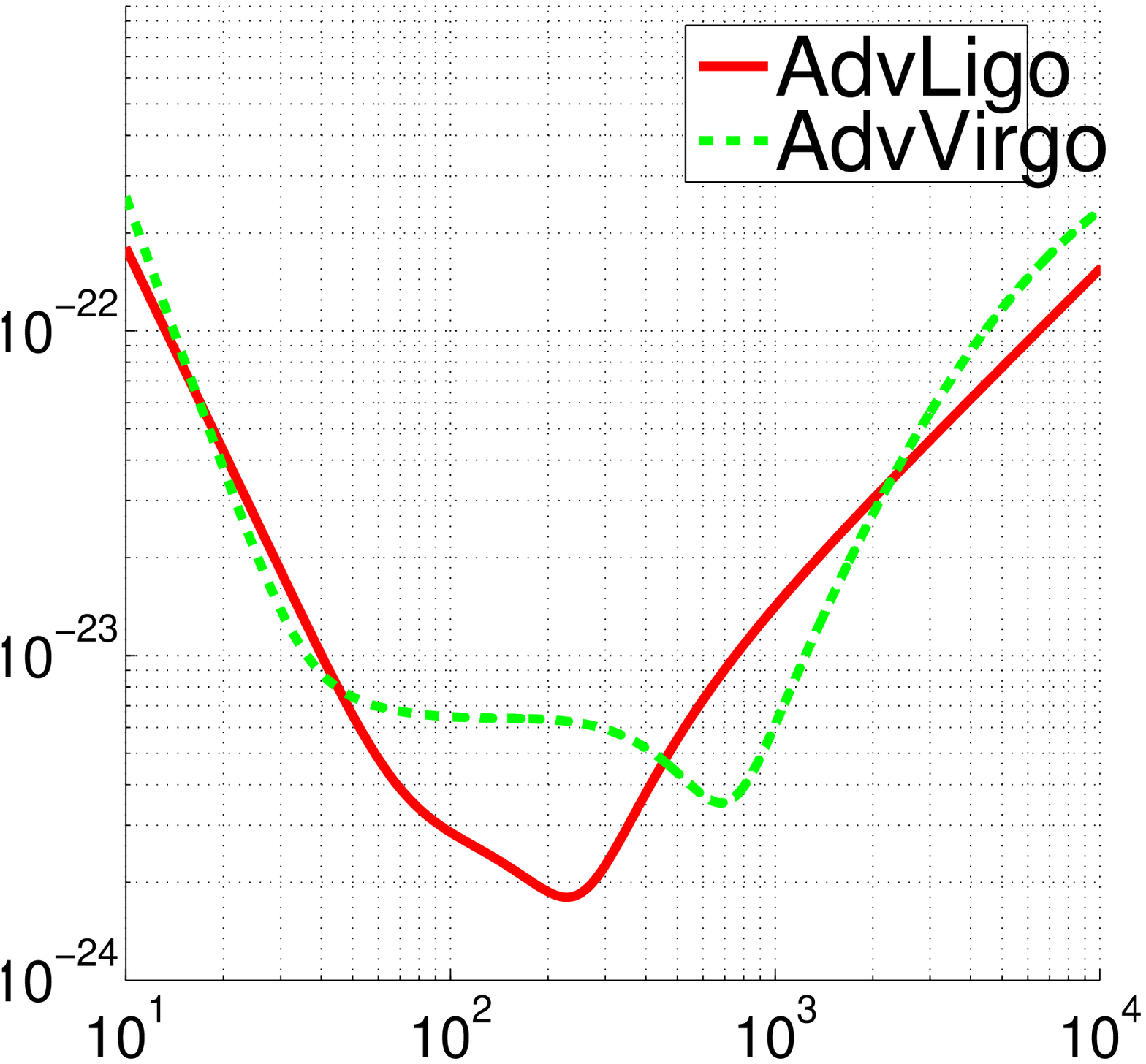} \\
&{ {\small $f$}}\quad 
\end{tabular}
\caption{\small Power spectral densities for AdvLigo (continuous line) and 
AdvVirgo (dashed line)}\label{Noises-IMR}
\end{figure}

\section{Results}\label{Section_Results}

To be able to compare our results with those of \cite{IMR-AjithApril2009}, we consider $\eta= 0.16, 0.2222, 0.25$ and $M=20,100,200 M_{\odot}$.
Tables 1 and 2 show the values of the first two orders of the covariance and bias for SNR=10. For each value of $\eta$ and $M_{\odot}$ the error estimation for $t_a$, $\phi_a$, total mass and symmetrized mass ratio $\eta$ are presented.

\begin{widetext}
\begin{figure}[htb]
 \begin{tabular}{*{2}{>{$} c <{$}|}||*{4}{>{$} c <{$} |}| *{4}{>{$} c <{$} |}|*{4}{>{$} c <{$} |}|}
&&\multicolumn{4}{c||}{\large $M=200 M_\odot$} & \multicolumn{4}{c||}{\large $M=100 M_\odot$} & \multicolumn{4}{c||}{\large $M=20 M_\odot$}\\
\hline\hline
& & \sigma[1] & \sigma[2] & b[1] & b[2] &
\sigma[1] &\sigma[2] & b[1] & b[2]&
\sigma[1] &\sigma[2] & b[1] & b[2]\\
 \hline\hline
\multirow{4}{5mm}{\begin{sideways}\parbox{15mm}{$\eta=0.25$}\end{sideways}} 
& \Delta t  & 9.47  & 12.5 & - 0.45 & - 1.04 & 3.54 & 2.92 & -0.10& 0.07 & 0.22 & 0.20
&-6.6\, 10^{-3}&-8.9\, 10^{-3} \\
\cline{2-14}
 &\Delta \phi & 40.7 & 59.7 & -7.85 & - 18.0 & 24.2 & 21.6 & -2.27 & - 0.98 & 9.98 & 11.0& - 0.42 & -0.57\\
\cline{2-14}
&\Delta M[\%] & 6.02 &9.60&0.82&1.87 & 2.61 & 2.12 & - 2.24\,10^{-4}& - 2.92\,10^{-2}& 1.38 & 1.44 & 5.34\,10^{-3} & 1.25\,10^{-2} \\\cline{2-14}
&\Delta \eta[\%]  & 11.3 &16.4 &0.81& 1.26&6.53&5.82&0.37& 0.58 & 2.58&2.88&0.04&3.64\,10^{-2} \\ \hline \hline
\multirow{4}{5mm}{\begin{sideways}\parbox{15mm}{$\eta=0.22$}\end{sideways}} 
& \Delta t  & 12.0  & 16.3 & -0.74& - 1.35& 4.33& 3.73& -0.16& 3.79\,10^{-2} & 0.23& 0.22
&-6.14\, 10^{-3}&-8.24\, 10^{-3} \\
\cline{2-14}
 &\Delta \phi  & 63.3 & 94.7 & - 9.69 & -21.6 & 36.5 & 33.9 & -2.70 & - 1.1 & 13.0 & 14.8& - 0.36 & -0.49\\
\cline{2-14}
&\Delta M[\%]  & 5.10 &8.21&0.74&1.93 & 2.62 & 1.88 & 9.63\,10^{-3}& - 3.57\,10^{-2}& 1.27 & 1.37 & 4.52\,10^{-3} & 0.70\,10^{-2} \\\cline{2-14}
&\Delta \eta [\%]  & 12.1 &18.0 &0.84& 1.77& 6.87&6.35&0.34& 0.51 & 2.36&2.72& 3.38\,10^{-2}&4.03\,10^{-2} \\ \hline \hline
\multirow{4}{5mm}{\begin{sideways}\parbox{15mm}{$\eta=0.16$}\end{sideways}} & 
\Delta t & 21.8  & 30.0 & -2.12 & -3.17 & 7.04 & 6.65 & -0.31 & -1.04 & 0.25 & 0.26
&-4.93\, 10^{-3}& - 5.79\, 10^{-3} \\
\cline{2-14}
 &\Delta \phi & 143 & 213 & - 18.0 & -33.2 & 77.0 & 77.7 & - 3.75 & - 2.26 & 19.2 & 22.4& - 3.76 & - 0.25\\
\cline{2-14}
&\Delta M[\%]  &3.07 & 4.21 & 0.48 &1.26 & 2.75 & 1.70 & 1.27 \,10^{-3}& - 4.84\,10^{-2}& 0.99 & 1.10 & 2.19\,10^{-3} & -1.23\,10^{-3} \\\cline{2-14}
&\Delta \eta [\%] & 14.2 & 20.8 & 0.85 & 2.37 & 7.53 & 7.58 & 0.37 & 0.52 & 1.84 & 2.15& 0.02 &3.57\,10^{-2} \\ \hline \hline
 \end{tabular}
\caption{The errors in an Advanced Ligo detector (table above) and Advanced Virgo (below). $\sigma[1]$ and $\sigma[2]$ are the first (the usual CRLB) and second order in the variance expansion; while $b[1]$ and $b[2]$ are the first and second order of the bias (see \ref{IMR-Sigma_series} and \ref{IMR-bias_series}). The time errors are in milliseconds, the phase errors are in radians, while the errors in the mass parameters are in percent. The SNR is equal to 10}
\end{figure}
\end{widetext}

\begin{widetext}
 \begin{tabular}{>{$} c <{$}| >{$} c <{$} || >{$} c <{$} | >{$} c <{$} | >{$} c <{$} | >{$} c <{$} ||>{$} c <{$} | 
>{$} c <{$} | >{$} c <{$} | >{$} c <{$} ||>{$} c <{$} | >{$} c <{$} | >{$} c <{$} | >{$} c <{$} ||}
&&\multicolumn{4}{c||}{\large $M=200 M_\odot$} & \multicolumn{4}{c||}{\large $M=100 M_\odot$} & \multicolumn{4}{c||}{\large $M=20 M_\odot$}\\
\hline\hline
&\tiny  &\tiny \sigma[1] & \tiny \sigma[2] &\tiny b[1] & \tiny b[2] &
\tiny \sigma[1] &\tiny \sigma[2] &\tiny b[1] &\tiny b[2]&
\tiny \sigma[1] &\tiny \sigma[2] &\tiny b[1] &\tiny b[2]\\
 \hline\hline
\multirow{4}{5mm}{\begin{sideways}\parbox{15mm}{$\eta=0.25$}\end{sideways}} & 
\Delta t & 10.8  & 11.0 & - 0.45 & - 0.17 & 5.69 & 6.30 & -0.15& 1.39 & 0.17 & 0.13
&-2.64\, 10^{-3}&-2.06\, 10^{-3} \\
\cline{2-14}
 &\Delta \phi & 42.7 & 49.9 & -8.04 & - 9.37 & 38.9 & 43.28 & -5.04 & 4.40 & 7.22 & 5.38& - 0.19 & -0.14\\
\cline{2-14}
&\Delta M[\%]  & 5.58 &7.07&0.55&0.64 & 3.85 & 3.89 & - 0.14 & - 1.25& 0.99 & 0.71 & - 7.04\,10^{-5} & 3.37\,10^{-3} \\\cline{2-14}
&\Delta \eta [\%] & 11.7 & 12.9 & 1.00 & 1.30 & 10.4 & 11.8 & 1.14 & 4.28 & 1.86 &1.40& 2.64\,10^{-2}  &8.05\,10^{-3} \\ \hline \hline
\multirow{4}{5mm}{\begin{sideways}\parbox{15mm}{$\eta=0.22$}\end{sideways}} 
& \Delta t  & 13.8  & 15.5 & -0.85& - 0.64& 6.47& 6.74& -0.19& 1.14 & 0.18& 0.14
&-2.35\, 10^{-3}&-1.82\, 10^{-3} \\
\cline{2-14}
 &\Delta \phi & 68.3& 84.56 & - 10.69 & -14.15 & 54.6 & 57.4 & -4.58 & 6.18 & 9.37 & 7.26 & - 0.16 & -0.12\\
\cline{2-14}
&\Delta M[\%]  & 4.92 & 6.46 & 0.56& 0.88 & 3.78 & 3.52 & -0.13 & - 0.93 & 0.91 & 0.68 & - 0.55 \,10^{-3} & 0.17 \,10^{-2} \\\cline{2-14}
&\Delta \eta [\%]  & 13.0 &15.4 &1.05& 1.66& 10.3&10.9&1.04& 3.26 & 1.70&1.33& 2.29\,10^{-2}& 9.89\,10^{-3} \\ \hline \hline
\multirow{4}{5mm}{\begin{sideways}\parbox{15mm}{$\eta=0.16$}\end{sideways}} & 
\Delta t  & 25.6  & 31.7 & -2.50 & -2.05 & 8.06 & 7.07 & -0.19 & 0.25 & 0.21 & 0.17
&-1.80\, 10^{-3}& - 1.16\, 10^{-3} \\
\cline{2-14}
 &\Delta \phi  & 162 & 217 & - 21.1 & -26.1 & 89.6 & 80.0 & - 2.54 &  2.04 & 13.9 & 11.1& - 8.36 \,10^{-2} & - 5.63\,10^{-2}\\
\cline{2-14}
&\Delta M[\%] & 3.50 & 3.97 & 0.39 & 0.77 & 3.23 & 2.31 & -0.10 & -0.24 & 0.72 & 0.55 & - 7.86\,10^{-4} & -4.74\,10^{-4} \\\cline{2-14}
&\Delta \eta [\%]  & 16.0 & 20.7 & 1.27 & 3.05 & 8.74 & 8.00 & 0.74 & 1.00 & 1.33 & 1.06 & 1.47\,10^{-2} &9.23\,10^{-3} \\ \hline \hline
 \end{tabular}
\end{widetext}

The CRLB consistently underestimate the error at this SNR.
The errors for different values of the SNR $\rho$ can be obtained 
by multiplying the first orders for $10/\rho$ and the second orders by $100/\rho^2$ (and multiplying for 10 and 100 gives $S_1^2$ and $S_2^2$). 
A necessary condition for the validity of the CRLB can also be obtained as: 
\begin{equation}
\frac{S_2^2}{S_1^2*\rho^2}<1
\end{equation}
where the $<$ sign can be replaced by $<<$ depending on the accuracy needs.
Using the data in the tables above, we can plot the errors against the SNR, for fixed values of the total mass $M$ and the mass ratio $\eta$. For example Fig. \ref{against_snr} shows the first order variance for the arrival time estimation, and the total variance (first plus second order) for an equal masses system of total mass $M=200 M_\odot$.
\begin{figure}
\begin{tabular}{ll} 
\multirow{1}{10mm}[20mm]{$\Delta t [s]$}&\includegraphics[width=4.3cm]{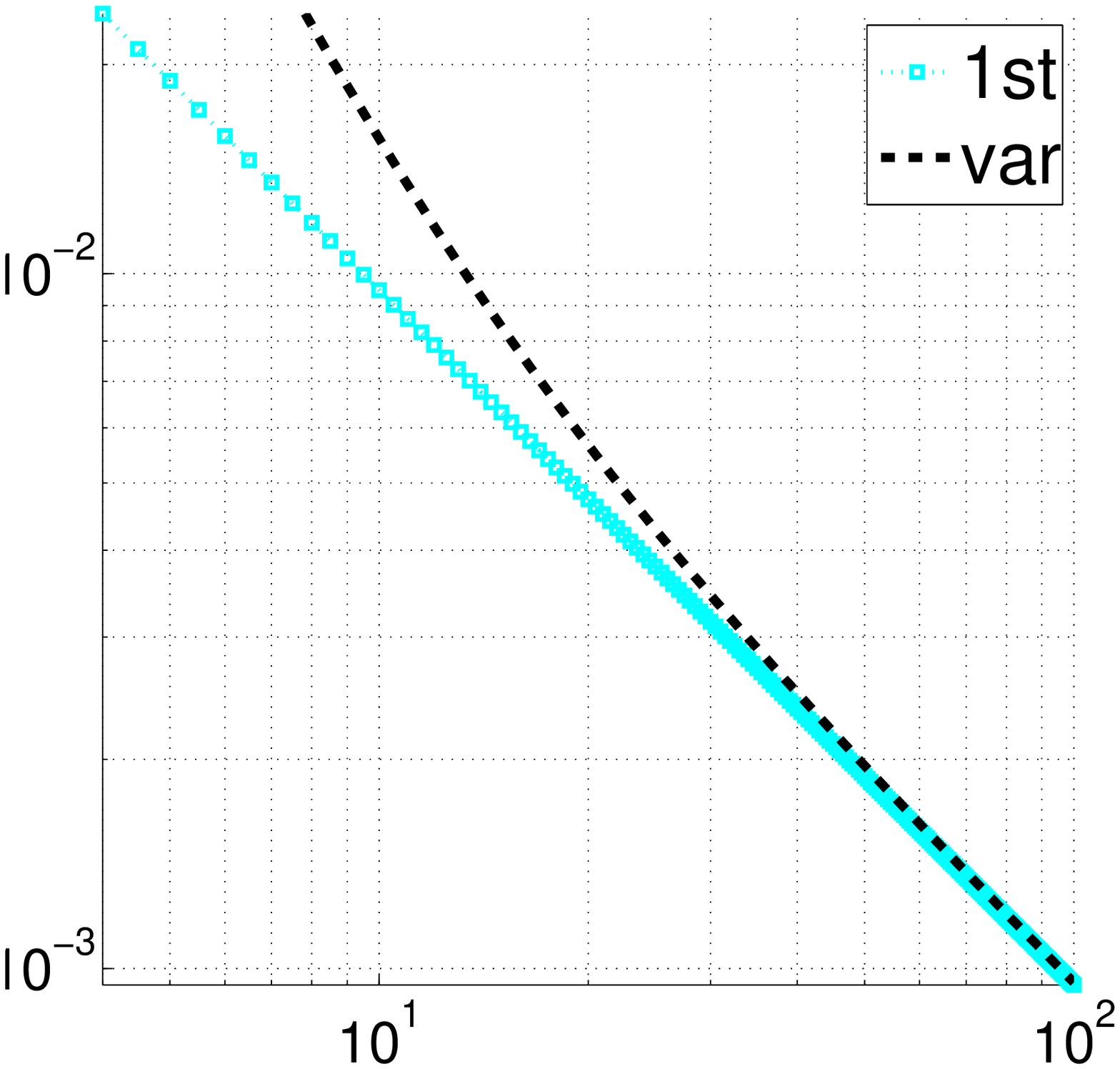}\\
\quad \qquad & \qquad\qquad\quad { {\small    $\rho$}}\quad 
\end{tabular}
\caption{\footnotesize The error in the estimation of the arrival time for a system with $M=200M_\odot$ and $\eta=0.25$. The dot-diamonds line is the first order variance (CRLB) while the dashed line is the total variance. The bias is not shown, being negligible.}\label{against_snr}
\end{figure}
It is clearly visible for which SNR the CRLB ceases to be faithful.

Another useful application is the calculation of the CRLB and second order variance for systems having mass and mass-ratio within a chosen range, building then a grid of results that clearly show for what systems the errors are smaller.
We have done that for $M=4 M_\odot .. 200 M_\odot$ and $\eta=0.10 .. 0.25$. In Fig. \ref{ContourM} we show a contour plot of the total error in the estimation of the system's mass (in percent), for systems having a fixed SNR of 10 and using the AdvLigo noise. 
Fig. \ref{ContourM} shows a not monotonic trend for the error: for exemple, a system having $\eta=.25$ will have an error of about $1\%$ if its mass is $4 M_\odot $. As the mass increases, keeping $\eta$ constant, (this is equivalent to scanning Fig. \ref{ContourM} from the lower right corner to the upper right one), there will be a local maximum of the error, due to the ``island'' on the right side of the plot, for $M\approx 55 M_\odot$, where the error reaches about the $5\%$, then the error goes down for a while, and begins to grow again, from about $M=100M_\odot$. On the other hand, if $\eta=0.16$, as the system's mass increases it won't find any ``island'' of great error, and no local maxima will be present. We will recover this behavior later (Fig. \ref{AgainstMAL}).

\begin{figure}
\begin{tabular}{ll} 
\multirow{1}{10mm}[20mm]{$\frac{M}{M_\odot}$}&\includegraphics[width=5cm]{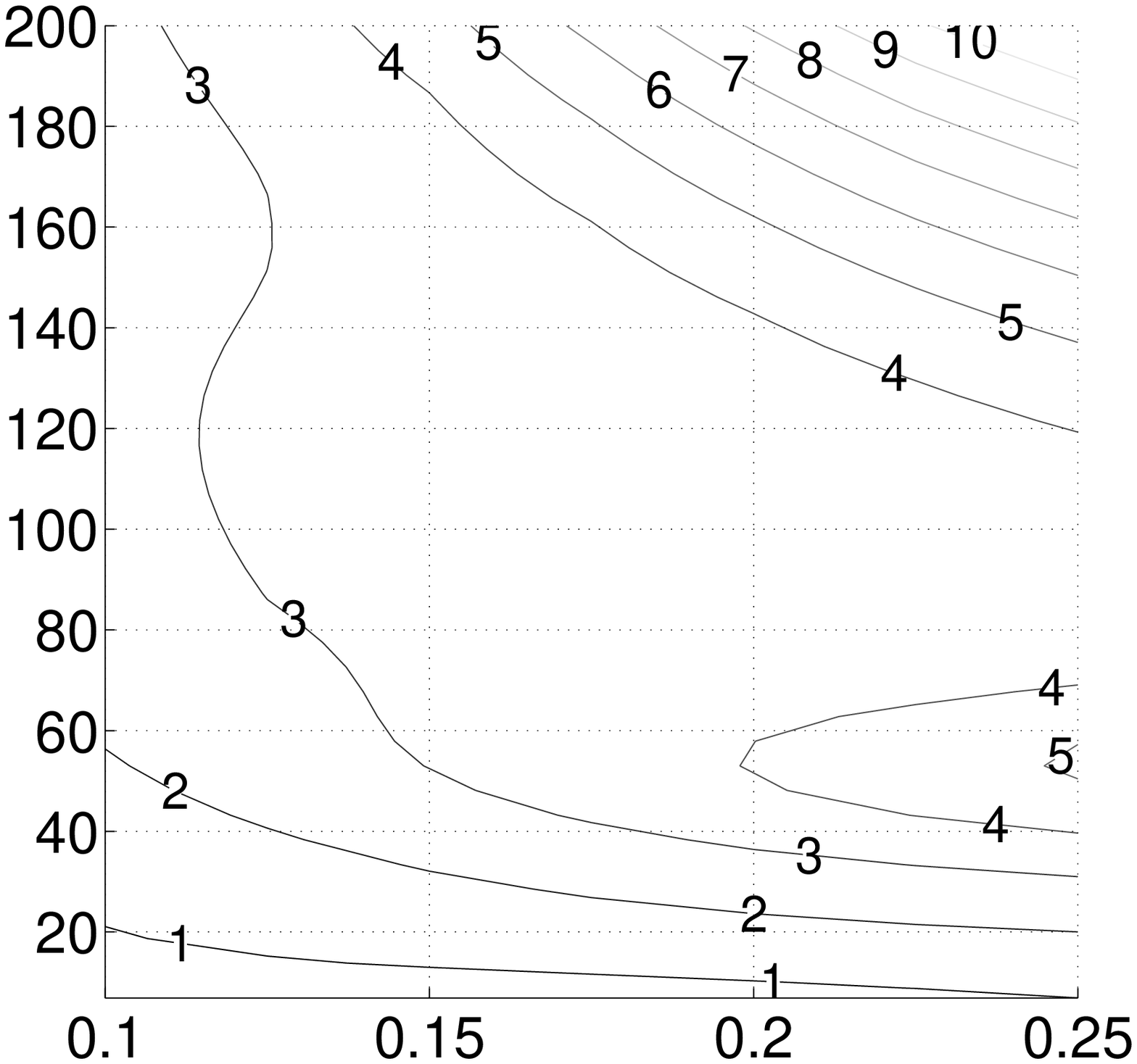}\\
\quad \qquad & \qquad\qquad\quad { {\small    $\eta$}}\quad 
\end{tabular}
\caption{\footnotesize The percent error on the total mass estimation (first plus second order) as a function of the total mass and symmetrized mass ratio. The systems have a fixed SNR, $\rho=10$, and the AdvLigo noise is used. This plot does not depend on the values of $t_a$ and $\phi_a$}\label{ContourM}
\end{figure}

We can use this kind of plots in another way. We calculate the ratio between the total variance (first plus second order) and the CRLB, and we let the SNR vary until the ratio goes under a chosen threshold \emph{for each point in the grid}. Let us for exemple plot $\frac{\sqrt{\sigma[1]^2_M +\sigma[2]^2_M}}{\sigma[1]_M}$ for $\rho=10$, this is shown in Fig. \ref{ratio_rho_10}. For highly symmetric massive systems the CRLB is nearly one half of the corrected error. If we want the ratio for the estimation of the total mass to be smaller than $1.05$ for each value of M and $\eta$ in the range considered, the SNR must be $\rho \geq 56.4$. We can perform the same kind of calculations for the other parameters. The biggest SNR we calculate in this way is the one required for the CRLB of the time parameter to attain a $5\%$ precision, $\rho=61.6$. Then we can say that for every binary system having mass and symmetric mass ratio in the range given above, a SNR of 61.6 assures that when using the CRLB for the errors' estimation our result will not differ more than $5\%$ from the corrected errors. 

It must be stressed that plots like those in Fig. \ref{ContourM} and Fig. \ref{ratio_rho_10} does not depend on the actual value of the arrival time and phase. The reason is the particular form of the signal, eq. (\ref{Wave_Fourier}), and the fact that $t_a$ and $\phi_a$ are only contained linearly in the phase of the signal. Let us for exemple consider the $(i,j)$ element of the CRLB, it depends on the real part of $h_i h_j^{*}$. Developing the derivations we have:

\begin{eqnarray}
\Re[ h_i h_j^{*}] &=& \Re \left[ \left(A_i + A\,(i {\psi}_i)\right)\,e^{i \psi}\,\left(A_j + A\,(-i {\psi}_j)\right)\,e^{- i \psi} \right] \nonumber\\
&=& A_i A_j + A^2 \psi_i\psi_j 
\end{eqnarray}

The arrival time and arrival phase will not be contained in terms like $A$ or $A_i$, as the amplitude does not depend on them. Neither they will be in terms like $\psi_a$, because they were contained linearly in $\psi$.
It is easy to see that the same kind of proof holds while calculating the $\upsilon_{\cdots}$, eqs. (\ref{IMR-vab}) to (\ref{IMR-vabcde}), and the optimal SNR, eq. (\ref{optimal_snr}).

\begin{figure}
\begin{tabular}{ll} 
\multirow{1}{10mm}[20mm]{$\frac{M}{M_\odot}$}&\includegraphics[width=5cm]{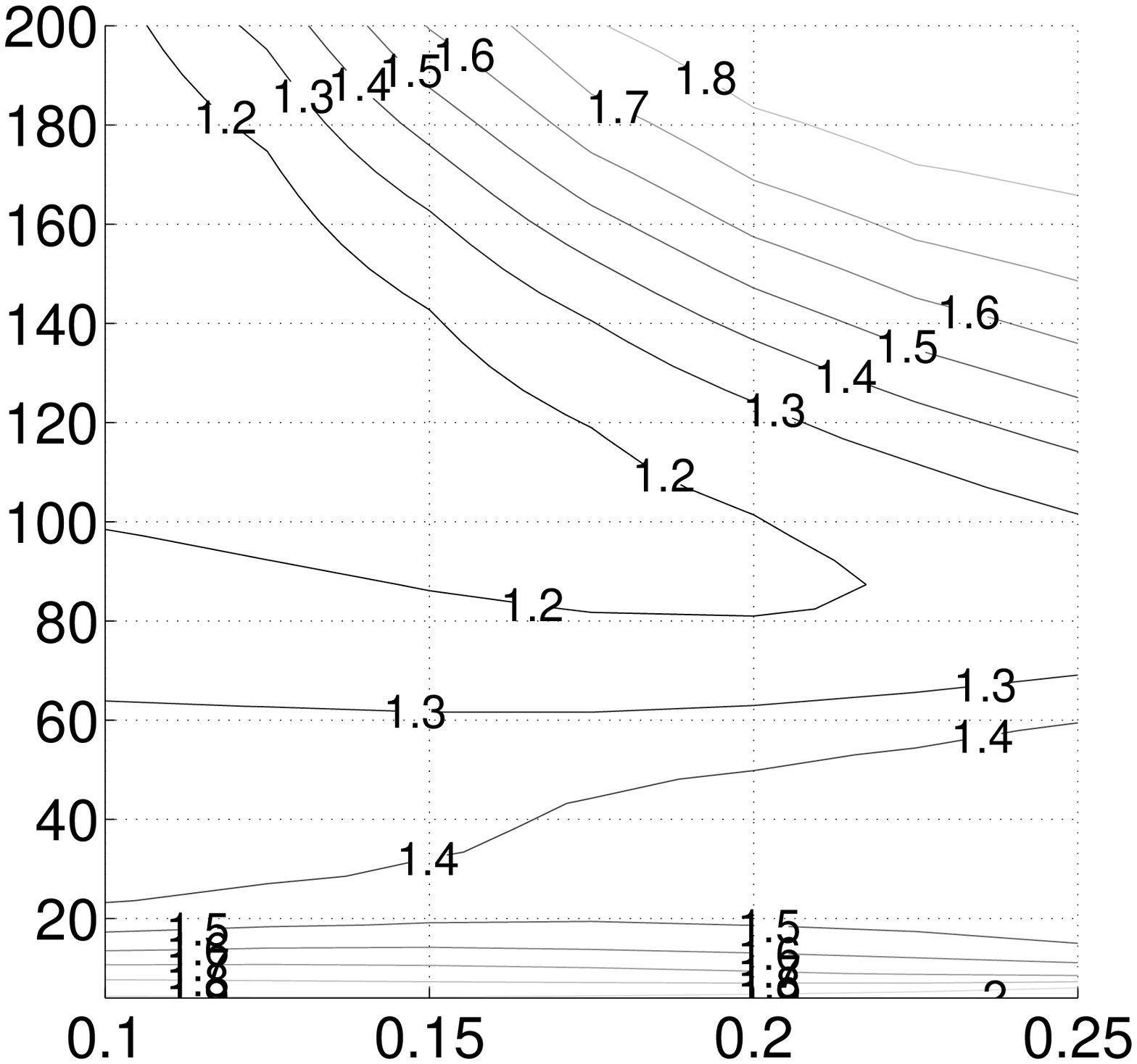}\\
\quad \qquad & \qquad\qquad\quad { {\small    $\eta$}}\quad 
\end{tabular}
\caption{\footnotesize The ratio between the total mass error (first plus second order) and the CRLB, as a function of the total mass and simmetrized mass ratio. The systems have a fixed SNR, $\rho=10$, and the AdvLigo noise is used. }\label{ratio_rho_10}
\end{figure}

As a general trend above SNR=20 the results are consistent with the results derived in \cite{IMR-AjithApril2009} with both Monte Carlo simulations and the CRLB. Lower SNRs result in higher uncertainties, and occasionally the Monte Carlo simulations are below the CRLB.
This behavior has already been observed in \cite{IMR-Cokelaer2008}, for inspiral signals, where the author shows that the inconsistencies between Monte Carlo and CRLB are a consequence of the restriction $\eta < 0.25$ in the
templates bank. By incorporating templates with $\eta > 0.25$, \cite{IMR-Cokelaer2008} obtains a good agreement between Monte Carlo and CRLB for 1.4 − 1.4 and 5 − 5 solar masses systems. However, discrepancies are still present in 10 − 10 solar masses systems. The author acknowledges that there is not a satisfactory explanation for this inconsistency. More efforts must be made to fully understand how the boundary $\eta = 0.25$ affects Monte Carlo
simulations and CRLB.

% This is not consistent to the statistical properties of the CRLB and might be due, as they state in \cite{IMR-AjithApril2009}, to the fact that the Monte Carlo simulations were exploring only the central peak of the probability distribution and applied restrictions to the values of $\eta$.

A comparison with our results that involves only the inspiral phase \cite{IMR-ZanolinVitale2010} indicates that by adding the merger and ringdown phases the errors decrease, but the necessary SNR (defined below) for the covariance to attain the CRLB can be up to a factor of two.
For example looking at Fig. 8 of \cite{IMR-ZanolinVitale2010} 3 of the 4  rows are comparable
(not the third one because  in the IMR we use the total mass while in \cite{IMR-ZanolinVitale2010} the chirp mass). The SNR necessary to attain the CRLB in the IMR signals at $M_{\odot}=20$  is between 9 and 11 while for inspiral phase waveforms is between 4 and 7.  The absolute values of the errors are however larger for inspiral signals.
In this case the contribution to the SNR of the inspiral phase is 99 percent
(see for example Fig. \ref{snr}  or its low $M_{\odot}$ blow up in Fig. \ref{snr_detail}).

\begin{figure}
\centering
 \begin{tabular}{ll}
\multirow{1}{10mm}[20mm]{ $\rho^2_{i} [\%]$} & \includegraphics[width=4.3 cm]{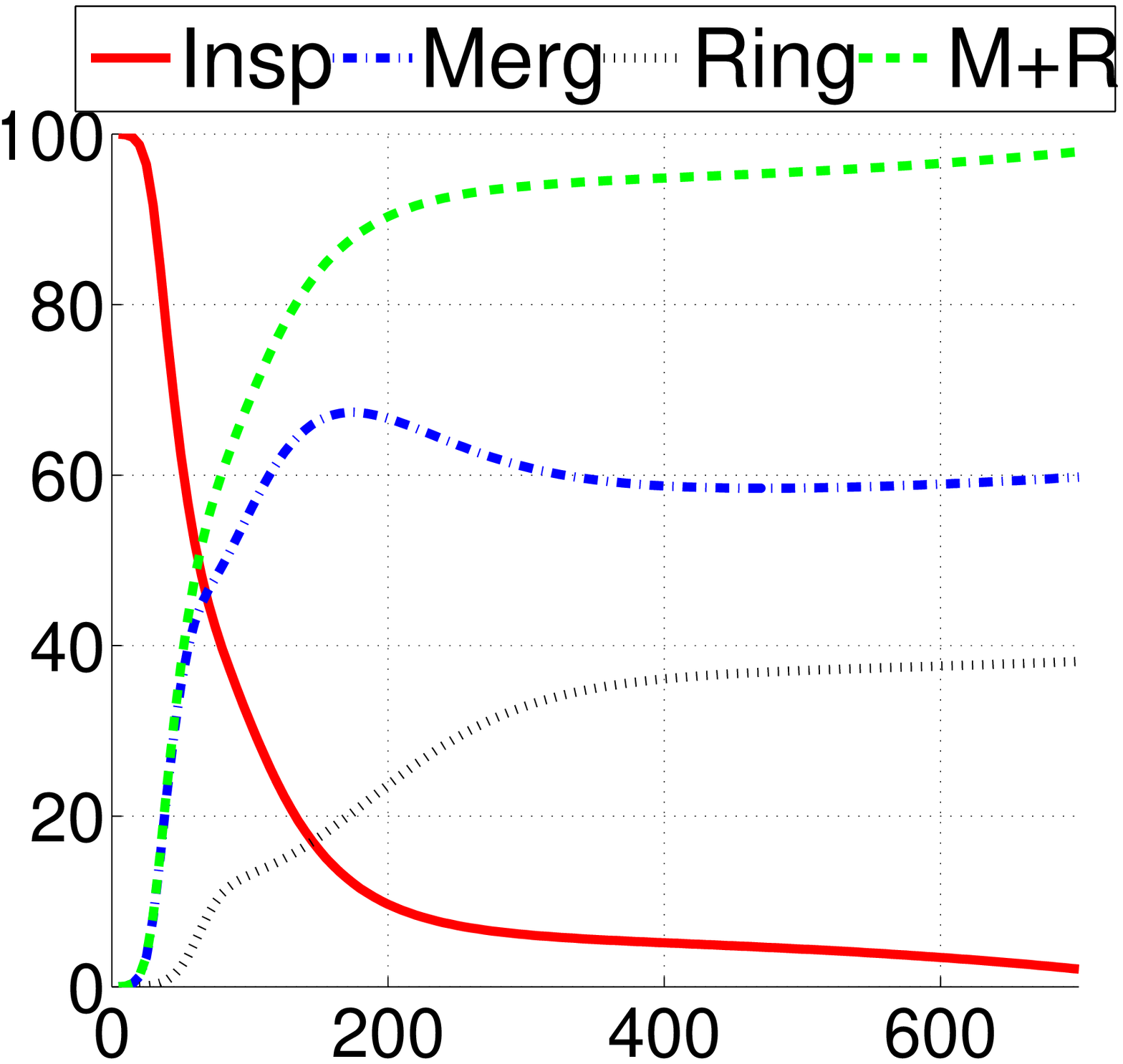} \\
\quad \qquad & \qquad\qquad\quad { {\small    $M$}}\quad 
\end{tabular}\caption{{\footnotesize The relative contribution of the different phases to the total squared SNR for an equal mass system of total mass M}\label{snr}}
\end{figure}

\begin{figure}
\centering
 \begin{tabular}{ll}
\multirow{1}{10mm}[20mm]{$\rho^2_{i} [\%]$} & \includegraphics[width=4.3 cm]{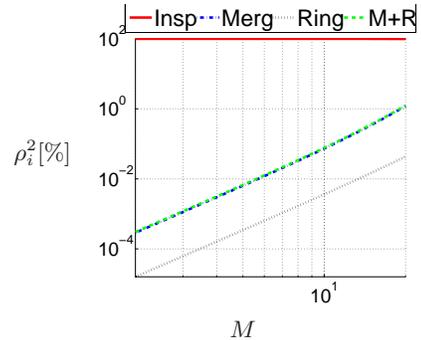} \\
\quad \qquad & \qquad\qquad\quad { {\small    $M$}}\quad 
\end{tabular}\caption{{\footnotesize The relative contribution of the different phases to the total squared SNR (low mass detail)}\label{snr_detail}}
\end{figure}

%In each panel we plot the total error, eq. (\ref{IMR-MSE}), the variance (first plus second order), the first order of the variance (ie the error as computed with the CRLB only), the second order of the variance, and the absolute value of the bias contribution (first plus second order of eq. \ref{IMR-bias_series}).
The bias does not play an important role, except for small SNRs ($<10$) or for high mass systems, for which the error in the total mass and arrival phase is seriously affected by the bias.
Unfortunately, while the use of the IMR allows for smaller values for the error estimation of the arrival time, total mass, and $\eta$, the same cannot be said of the arrival phase, for which the inclusion of the merger and ringdown phase seems to degrade the estimation, so that the error of the arrival phase estimation is in general higher than $2\pi$, indicating that the arrival phase is unpredictable. 
In \cite{IMR-Arun2005} the error for the arrival phase was estimated using the inspiral 3.5 PN wave, obtaining a value of $\Delta \phi=1.16 \;rad$ for a system of $M=20M_\odot$ at an SNR $\rho=10$ using the AdvLigo noise. For the same system using the IMR wave we obtain $\Delta \phi=14.8\; rad$ (9.98 rad considering the CRLB only). An estimation of the arrival phase error was not given in \cite{IMR-AjithApril2009}, and so a direct comparison is not possible. 
Since they are consistently above $2\pi$, the errors on $\phi$ are not included in the plots. Let us stress that the large errors on the arrival phase does not imply that we cannot believe the estimations we have for the other parameters. The reason is that a correlation coefficient different from zero, say close to +1, tells us that an overestimation of the parameter x comes together with an overestimation of the parameter y; but it doesn’t tell  anything abut the relative magnitude of the errors. 
It is interesting to plot the errors values against the total mass of the system, for a fixed value of the SNR.
Fig. \ref{AgainstMAL} shows these plots for a mass range from $4 M_{\odot}$ up to $500 M_{\odot}$, and an SNR $\rho=10$, using the AdvLigo noise. Fig. \ref{AgainstMAV} does the same with the AdvVirgo noise.
The inclusion of the second order variance and bias has visible consequences on the errors for large mass systems, for which the corrected error can be much larger than the CRLB.
The plots show an oscillatory character of the bias, due mainly to the behavior of $b[2]$.
For example  we plot in Fig. \ref{PlotBias} the bias orders on the total mass estimation against the system's mass, for $\eta=0.2222$. 

%\begin{equation}
%\clearpage

\begin{figure}
\centering
 \begin{tabular}{ll}
\multirow{1}{10mm}[20mm]{\small $\Delta t [s]$} & \includegraphics[width=4.0 cm]{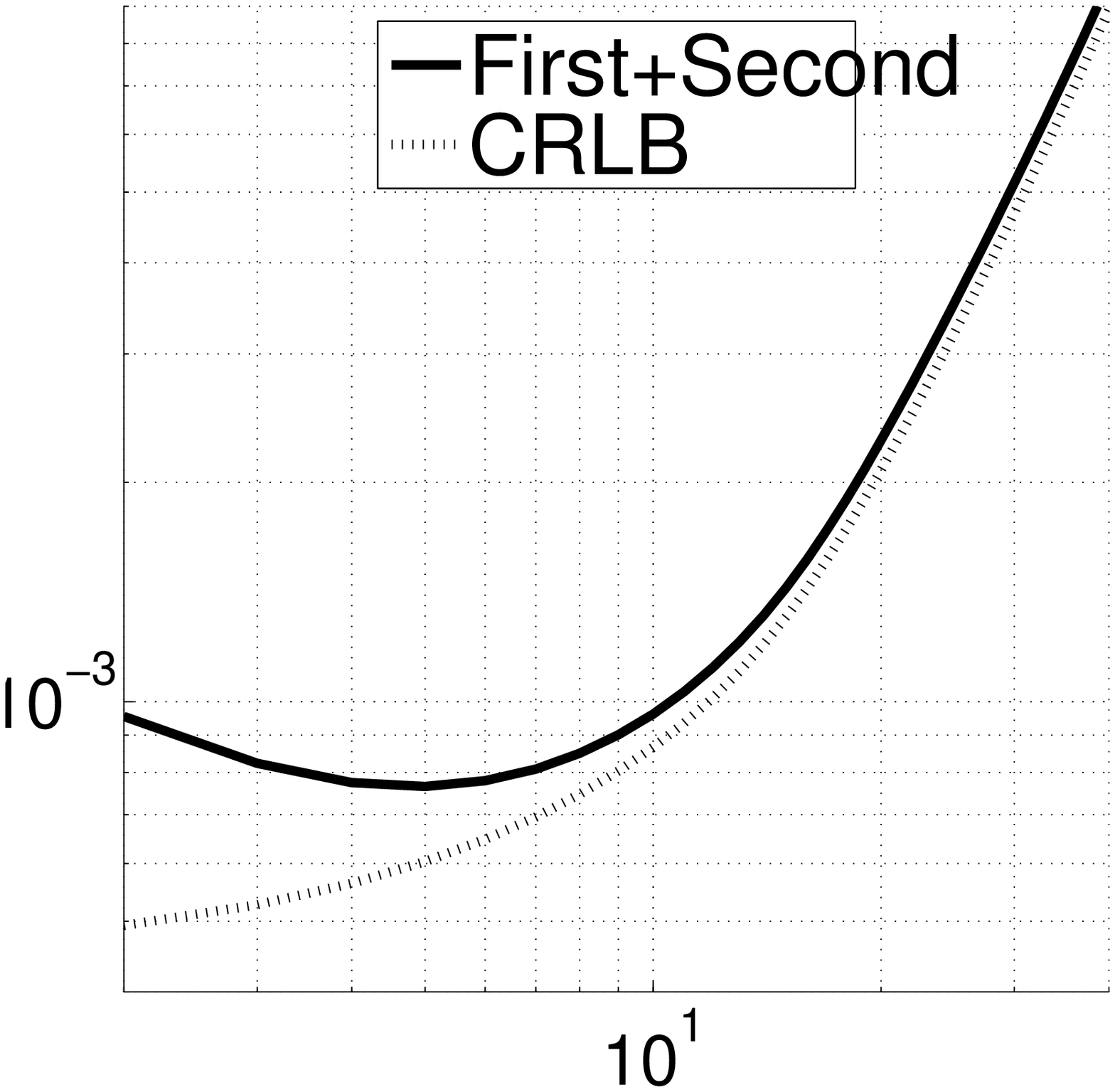} \\
\quad \qquad & \qquad\qquad\qquad { {\small    $\rho$}}\quad 
\end{tabular}\caption{{\footnotesize Minima in the error with respect to the mass can appear
 because of the contribution of the second order. In the plot the timing error is evaluated only
for the inspiral signal of $\eta=0.25$ at a fixed SNR, $\rho=10$. }\label{Plotminima}}
\end{figure}

The second order covariance and the bias also seem to reinforce the minimum 
of the errors for $M~100M_{\odot}$. It is important to notice that a local minimum would be present also with the inspiral phase only. For example in Fig. \ref{Plotminima} we show the error in the timing calculated for a 1.4-1.4 $M_{\odot}$ binary system where only the inspiral phase is used (3.5 Post-Newtonian waveform, \cite{IMR-Arun2005}), and a minimum is visible once the second order is included. To verify our new predictions in the errors for very large masses, numerical simulations or direct applications of parameter estimation pipelines are needed. Fig. 9 of \cite{IMR-AjithApril2009} shows how BHs merger can have very high SNRs, even of 100, at 100 megaparsecs for advanced LIGO. At these SNRs our corrections would not be important. However, in order to have useful detection rates we would need to rely on sources up to a gigaparsec. In fact, in \cite{Abadie2010} a realistic rate for BH merger is given as 0.4 per million years in an equivalent Milky Way galaxy. Equation 5 in \cite{Abadie2010} also gives an approximation of the number of equivalent Milky way galaxies. Combining these two observations suggests that a realistic rate for BH mergers within a gigaparsec is about 1.6 per year.
\begin{widetext}
 \begin{figure}[htb]
 \begin{tabular}{llllll}
\hskip -.2cm \multirow{1}{6mm}[20mm]{\small $ \Delta t [\mbox{s}]$} & \includegraphics[width=4.0 cm , height=3.4 cm]{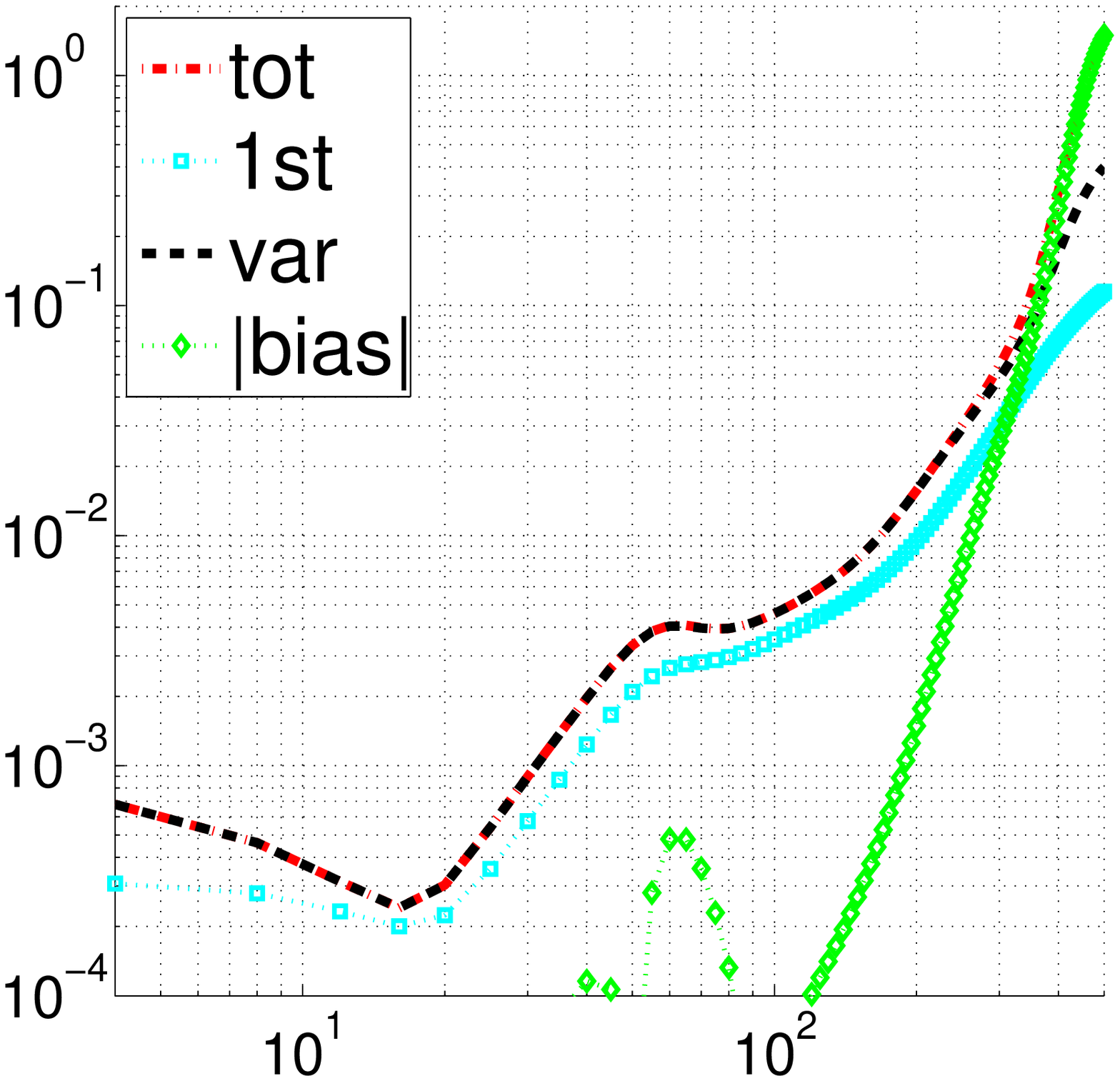} & \hskip -.2cm \multirow{1}{9mm}[20mm]{\small $ \frac{\Delta M}{M} [\mbox{\%}]$} & \includegraphics[width=4.0 cm , height=3.4 cm]{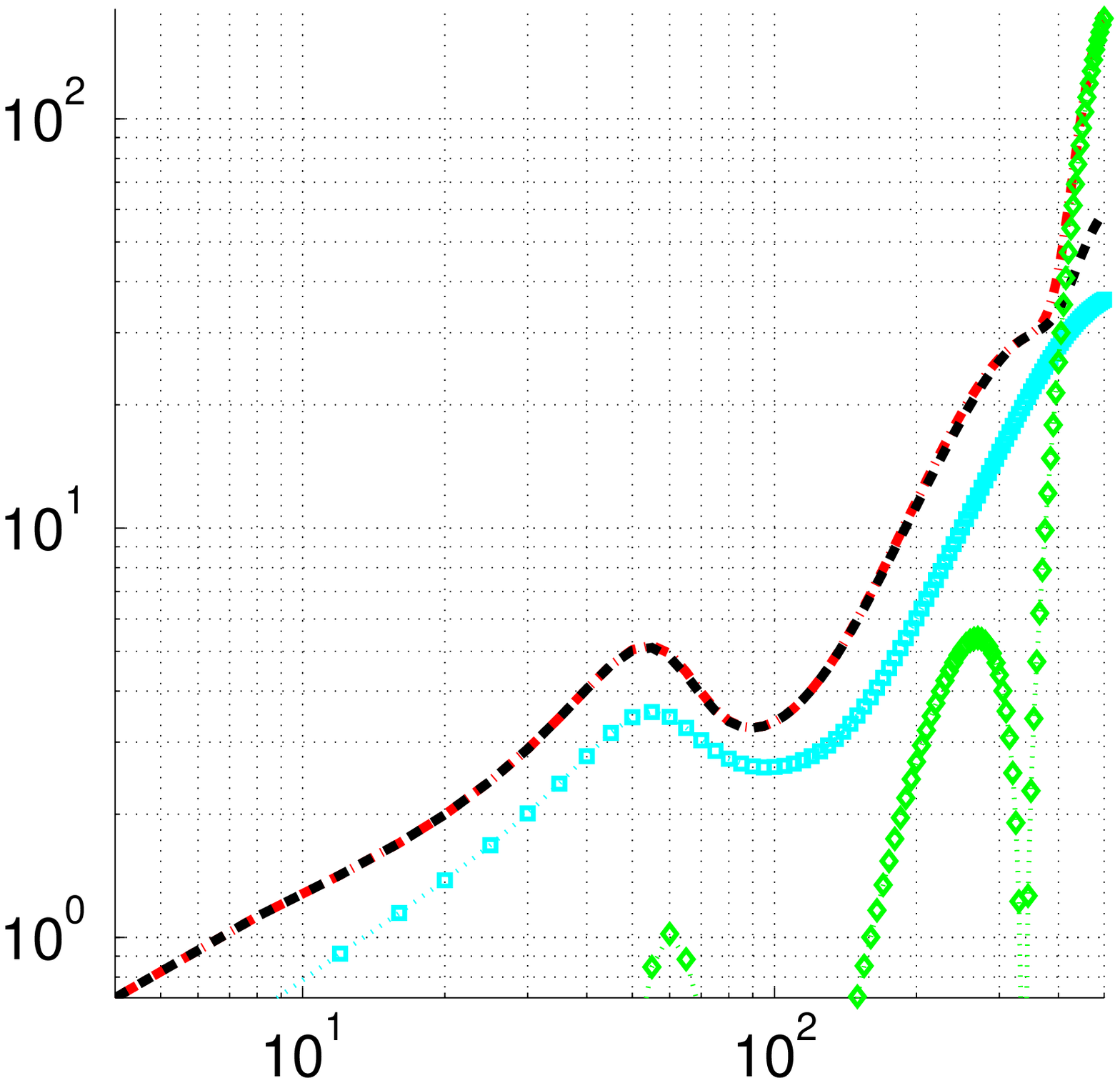} & \hskip -.2cm \multirow{1}{7.5mm}[20mm]{\small $ \frac{\Delta \eta}{\eta} [\mbox{\%}]$} & \includegraphics[width=4.0 cm,  height=3.4 cm]{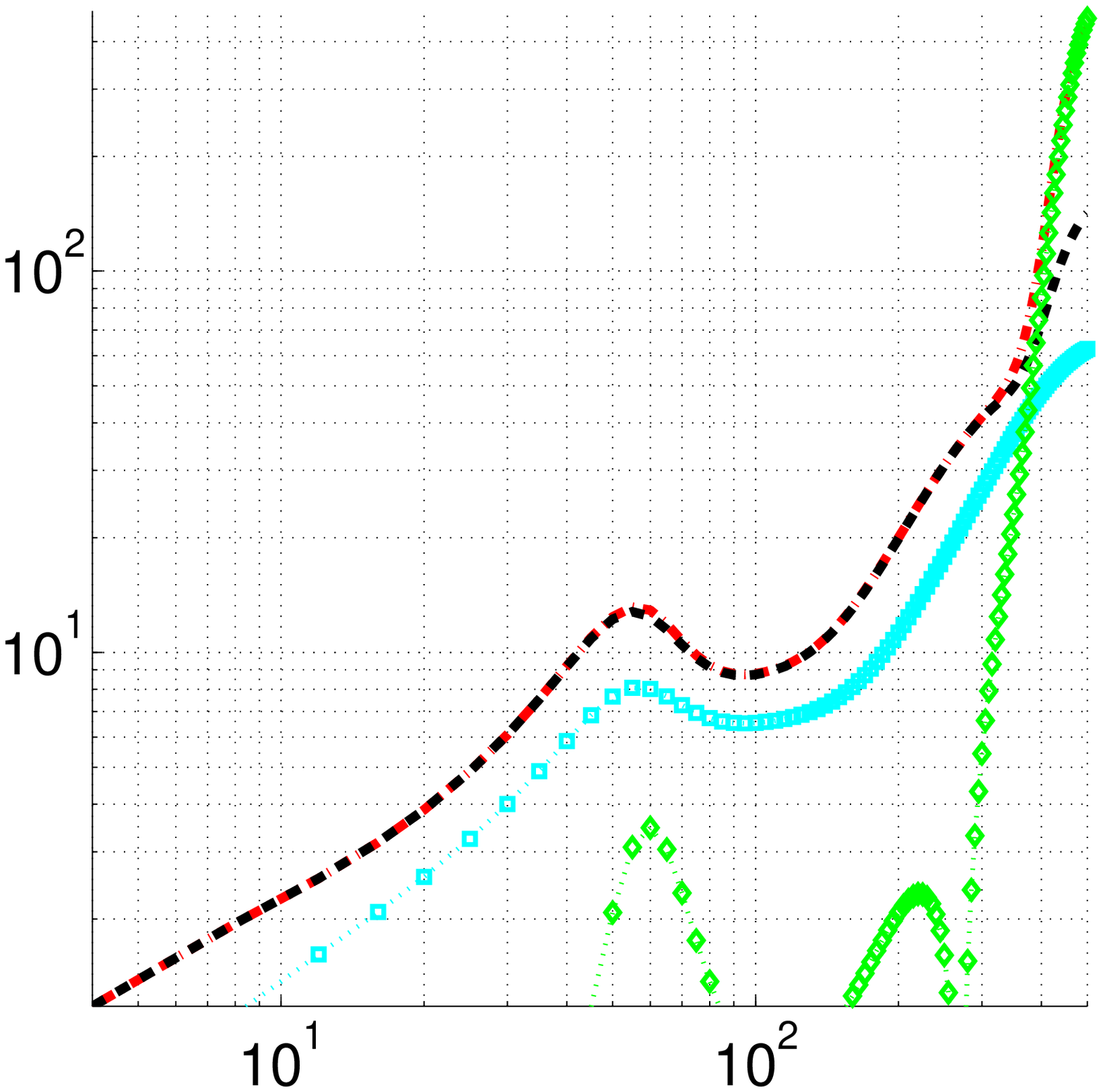} \\
\hskip -.2cm \multirow{1}{6mm}[20mm]{\small $ \Delta t [\mbox{s}]$} & \includegraphics[width=4.0 cm , height=3.4 cm]{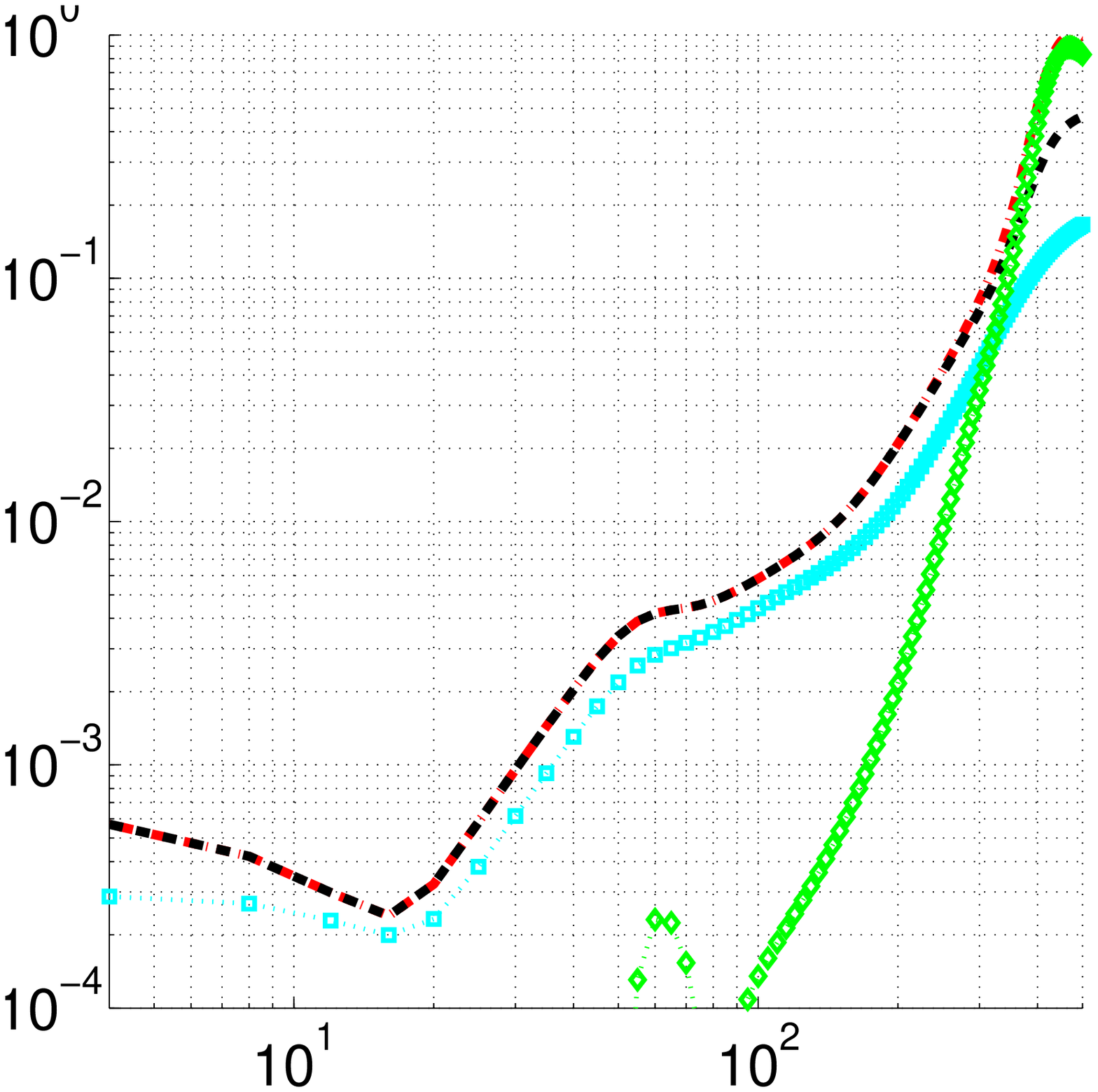} & \hskip -.2cm \multirow{1}{9mm}[20mm]{\small $ \frac{\Delta M}{M} [\mbox{\%}]$} & \includegraphics[width=4.0 cm , height=3.4 cm]{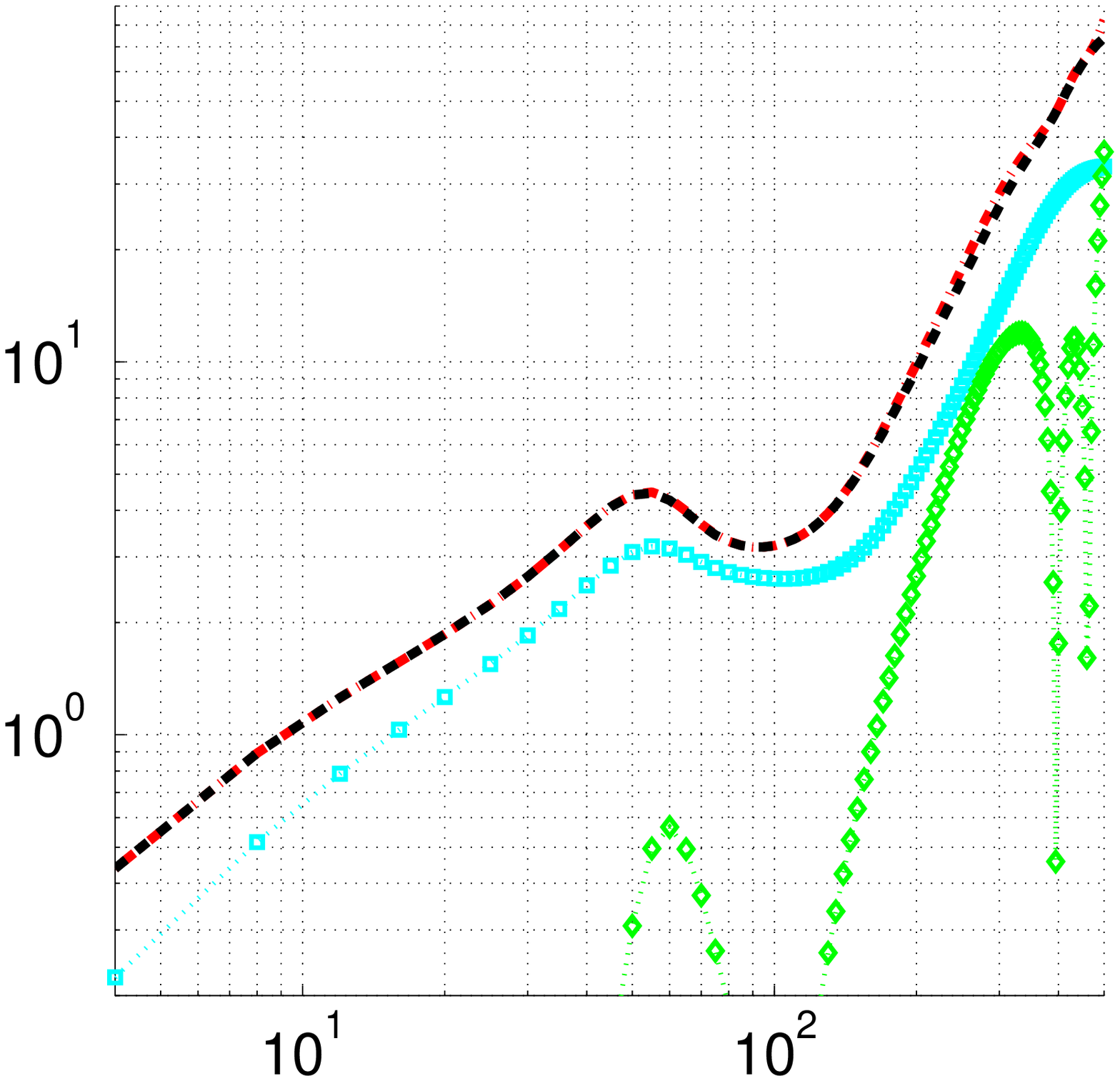} & \hskip -.2cm \multirow{1}{7.5mm}[20mm]{\small $ \frac{\Delta \eta}{\eta} [\mbox{\%}]$} & \includegraphics[width=4.0 cm,  height=3.4 cm]{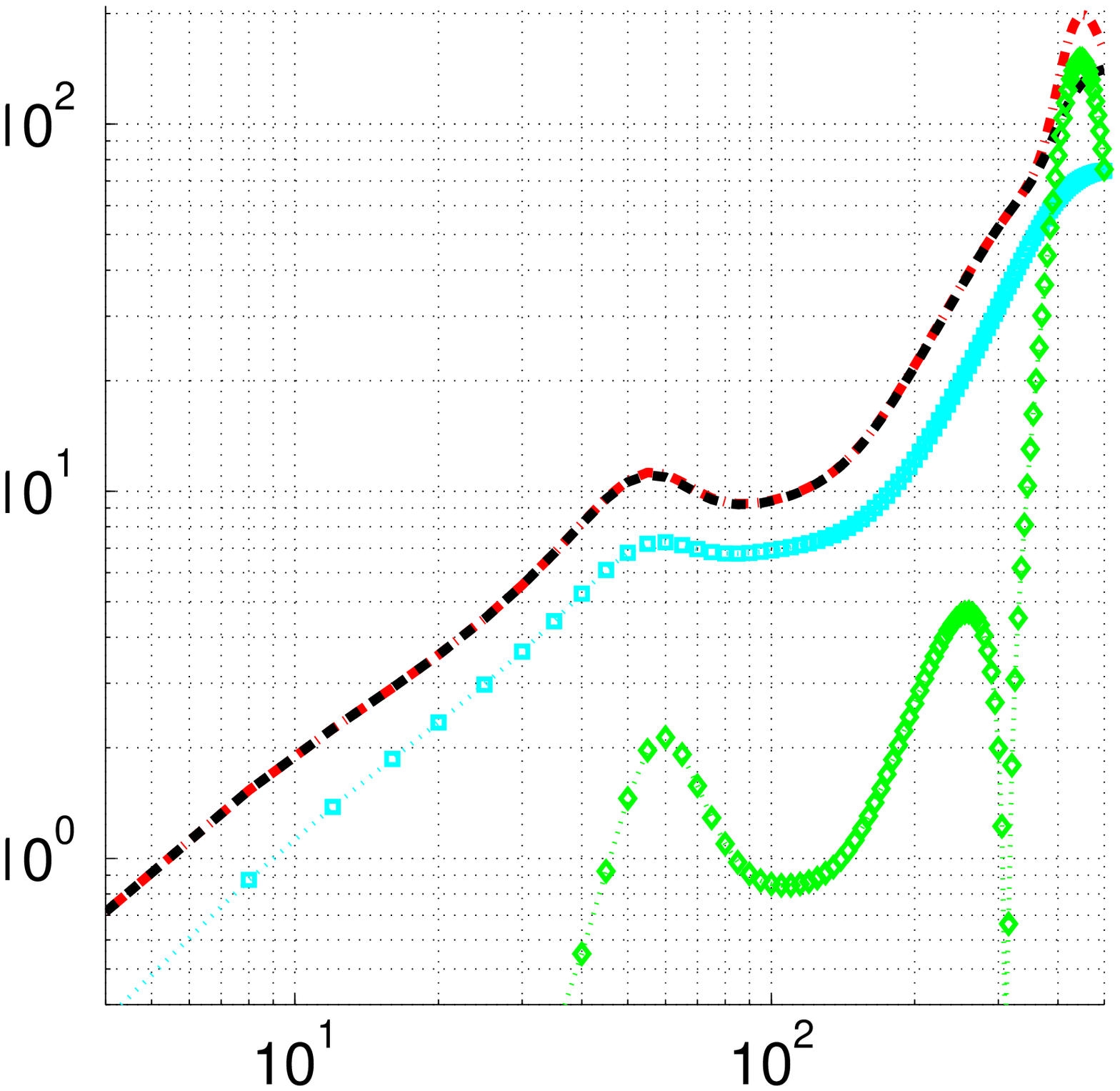} \\
\hskip -.2cm \multirow{1}{6mm}[20mm]{\small $ \Delta t [\mbox{s}]$} & \includegraphics[width=4.0 cm , height=3.4 cm]{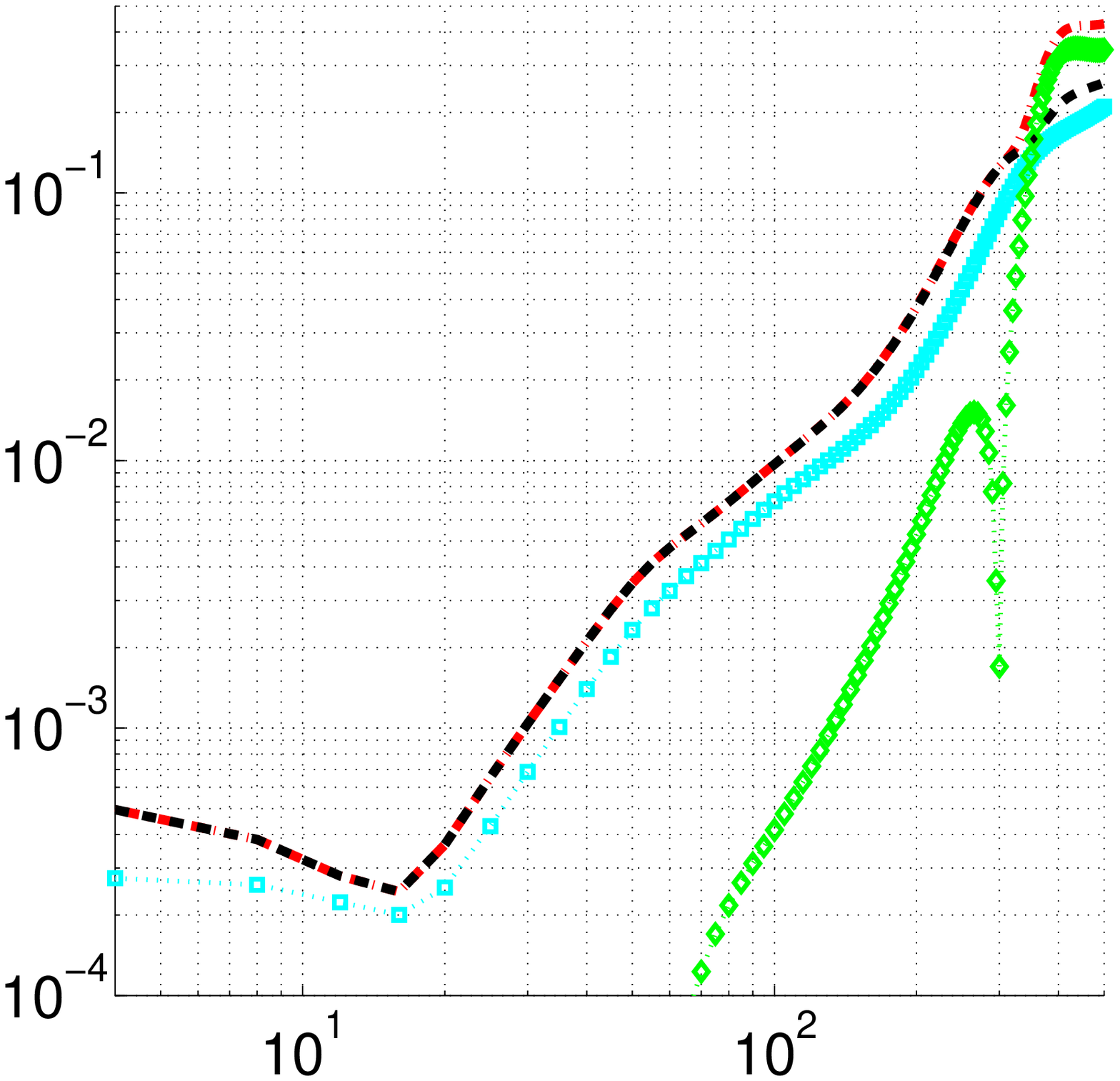} & \hskip -.2cm \multirow{1}{9mm}[20mm]{\small $ \frac{\Delta M}{M} [\mbox{\%}]$} & \includegraphics[width=4.0 cm , height=3.4 cm]{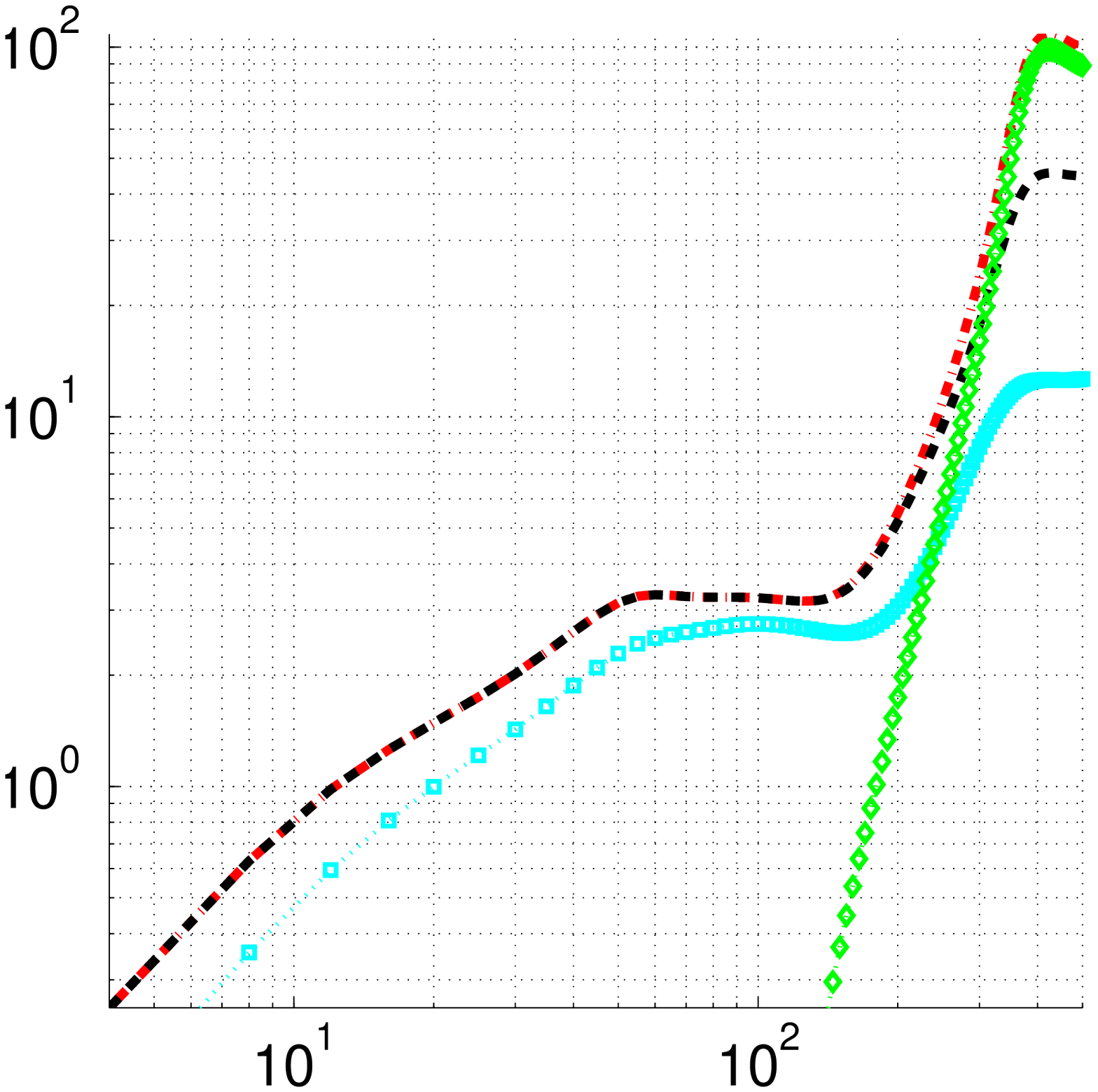} & \hskip -.2cm \multirow{1}{7.5mm}[20mm]{\small $ \frac{\Delta \eta}{\eta} [\mbox{\%}]$} & \includegraphics[width=4.0 cm,  height=3.4 cm]{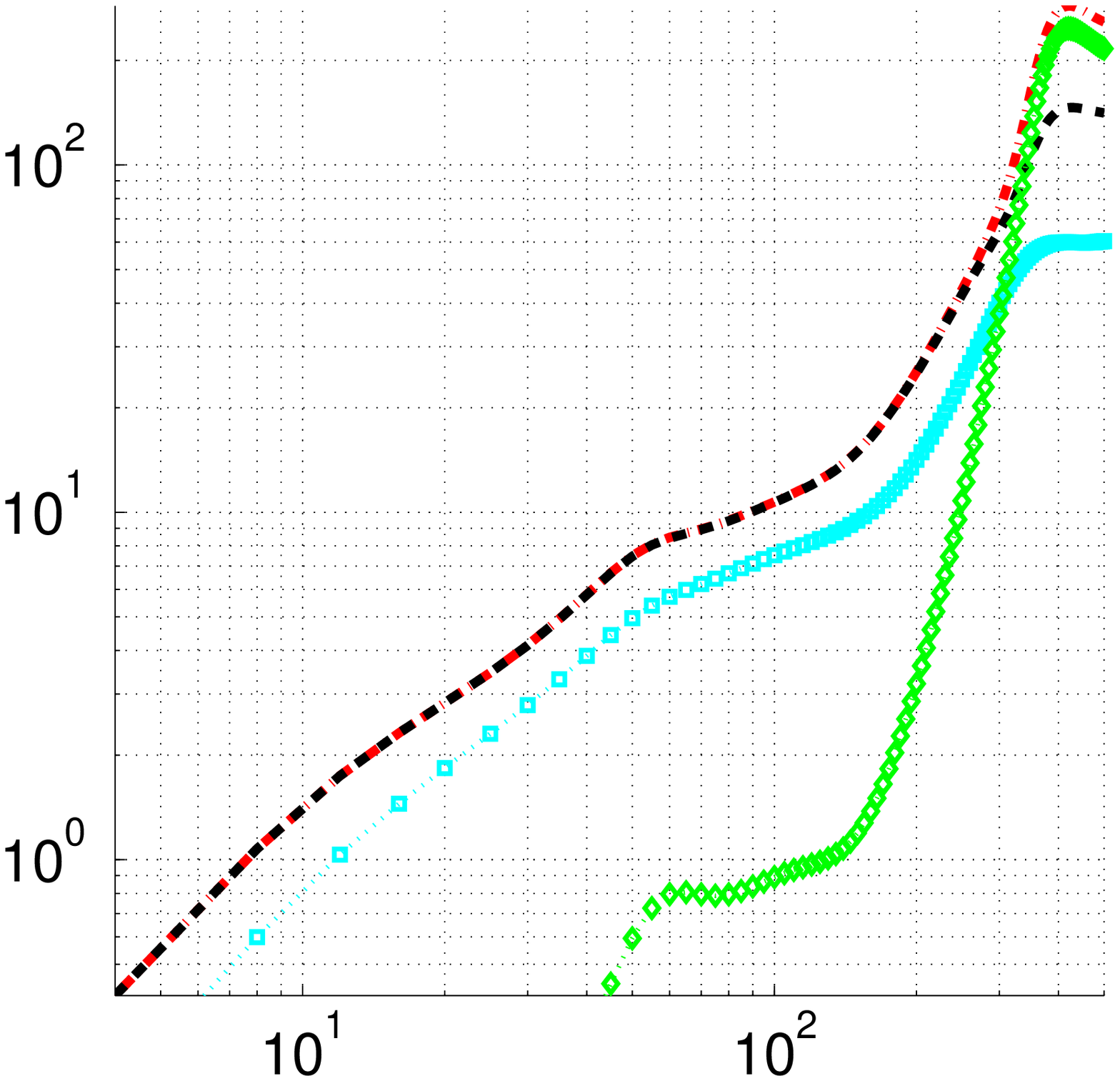} \\
&\hskip 10mm { {\small $M$}}\quad & &\hskip 10mm{ {\small $M$}}\quad & &\hskip 10mm{ {\small $M$}}\quad 
\end{tabular}
\caption{ \small(Color Online) The errors plotted against the total mass, using the AdvLigo noise, for a fixed value of $\rho=10$, and with $\eta=0.25$ (top), $\eta=0.2222$ (middle) and $\eta=0.16$ (bottom). The ``tot'' dot-dashed line represents the MSE (variance plus bias); the ``var'' dashed line the variance (first plus second order); the dot-squares line is the fist order variance  (CRLB); the dot-diamonds line, finally, is the absolute value of the bias (first plus second order). The ``tot'' and ``var'' lines are nearly superimposed, except for very high masses}\label{AgainstMAL}
\end{figure}
\end{widetext}
$$\\$$$$
$$\\
\newpage
\begin{widetext}
\begin{figure}[htb]
 \begin{tabular}{llllll}
\hskip -.2cm \multirow{1}{6mm}[20mm]{\small $ \Delta t [\mbox{s}]$} & \includegraphics[width=4.0 cm , height=3.4 cm]{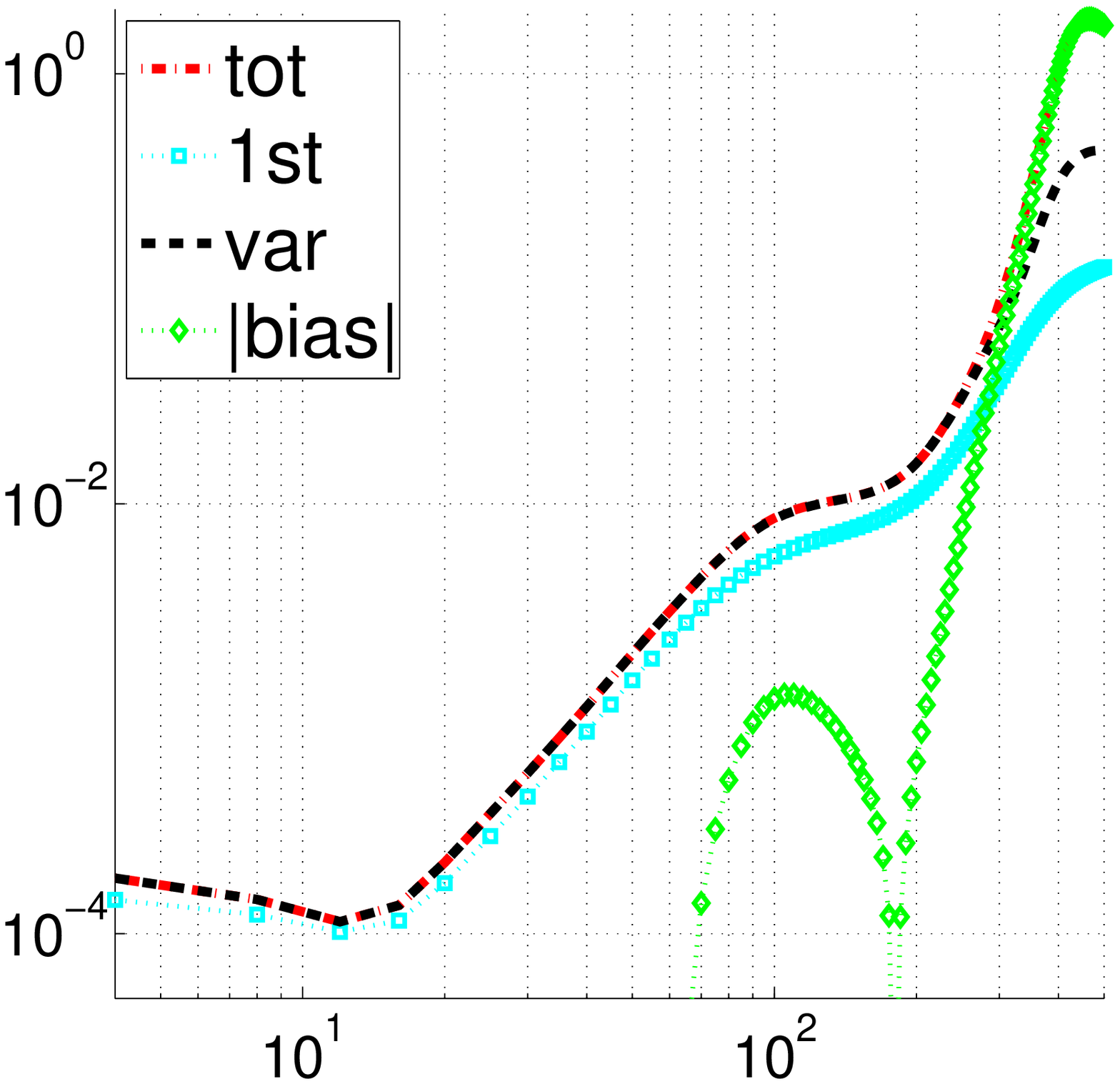} & \hskip -.2cm \multirow{1}{9mm}[20mm]{\small $ \frac{\Delta M}{M} [\mbox{\%}]$} & \includegraphics[width=4.0 cm , height=3.4 cm]{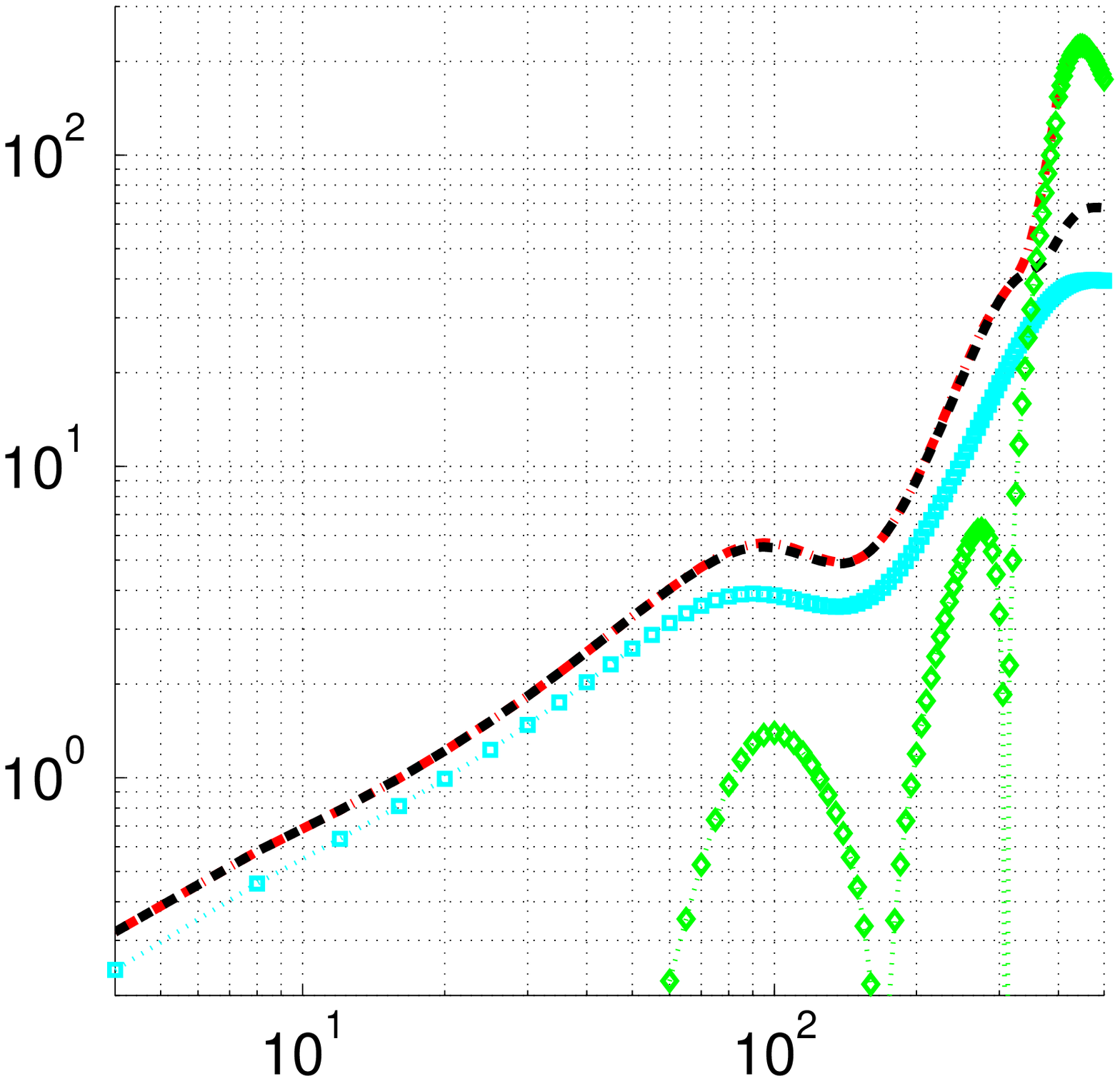} & \hskip -.2cm \multirow{1}{7.5mm}[20mm]{\small $ \frac{\Delta \eta}{\eta} [\mbox{\%}]$} & \includegraphics[width=4.0 cm,  height=3.4 cm]{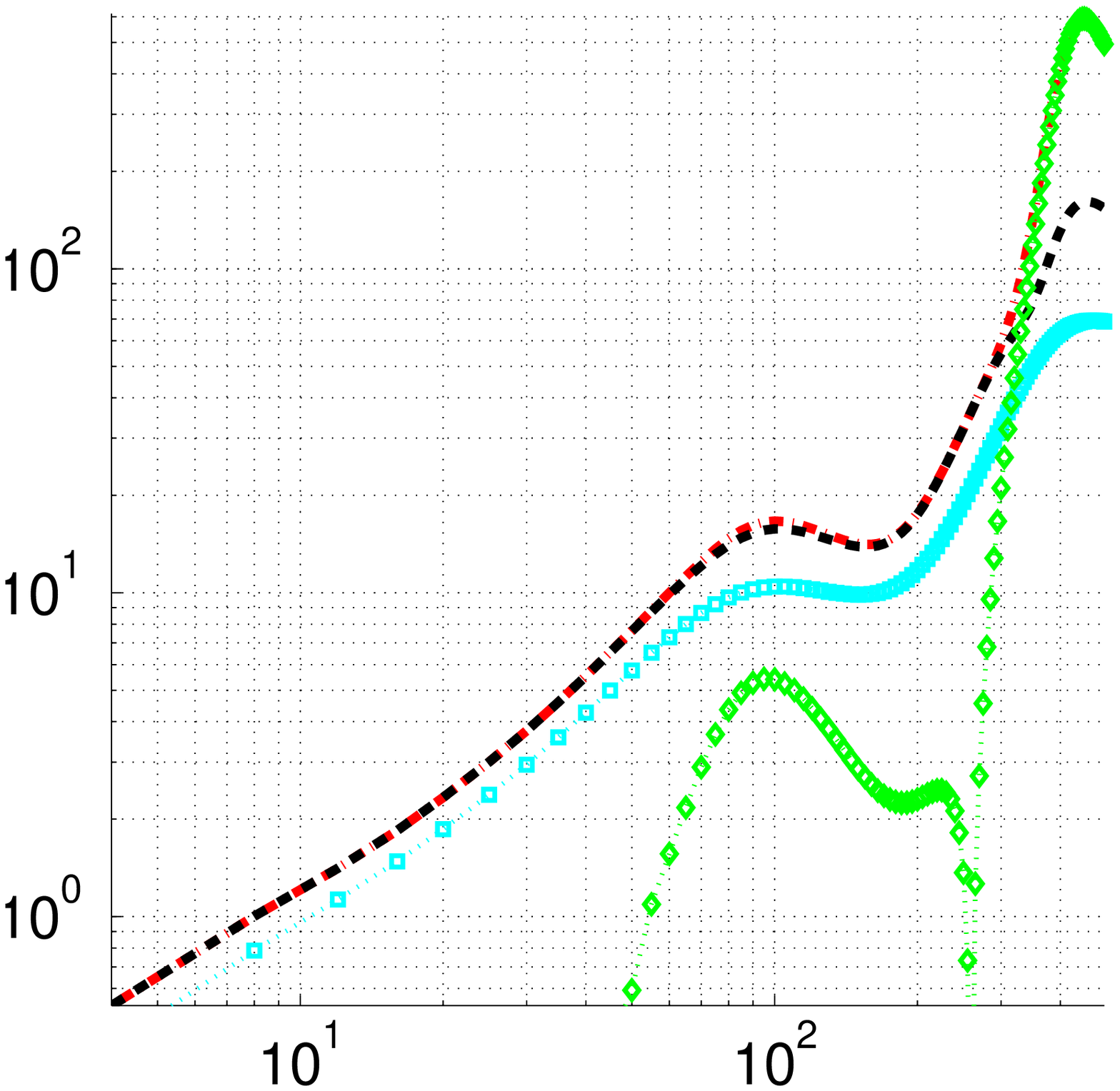} \\
\hskip -.2cm \multirow{1}{6mm}[20mm]{\small $ \Delta t [\mbox{s}]$} & \includegraphics[width=4.0 cm , height=3.4 cm]{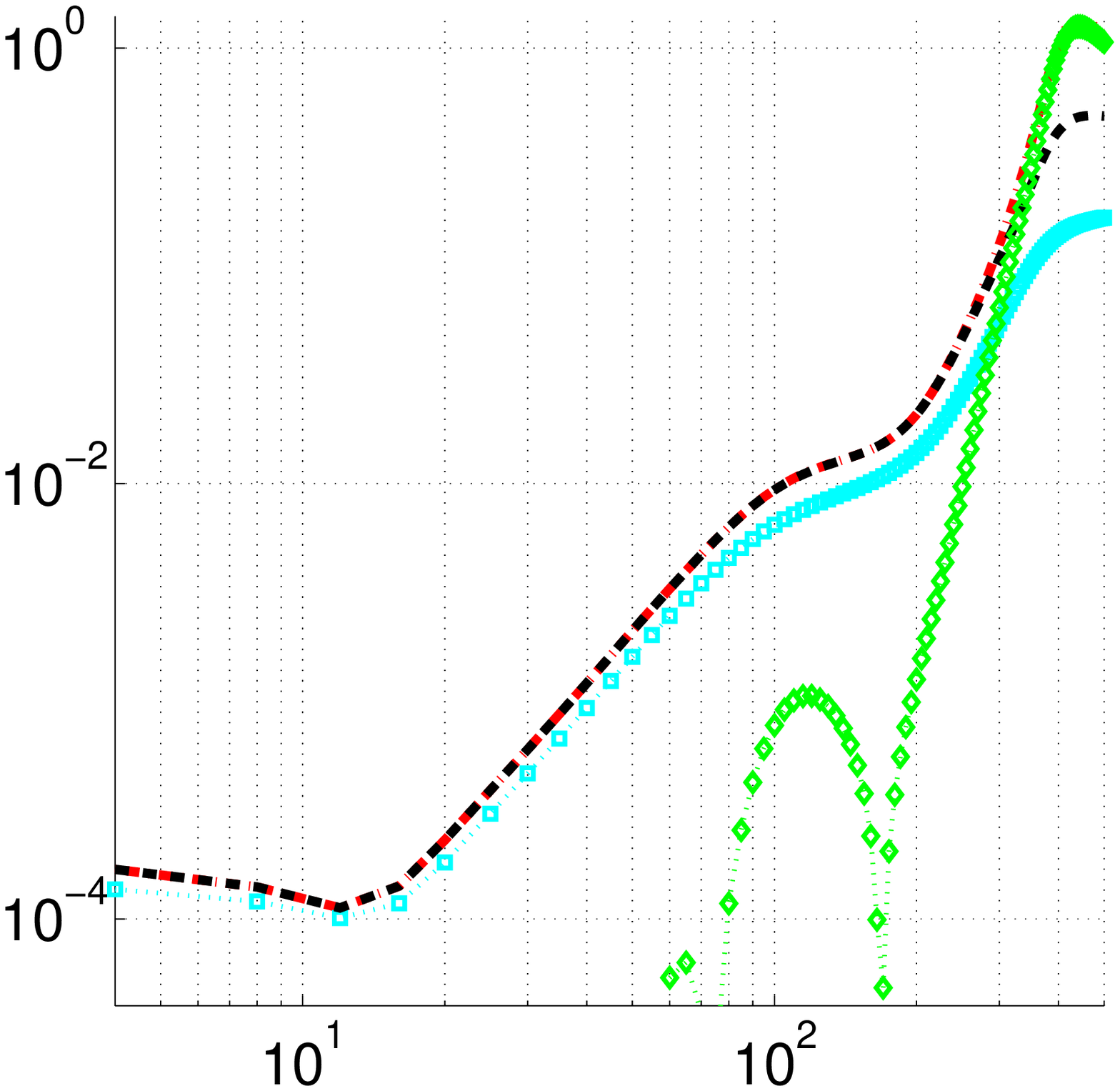} & \hskip -.2cm \multirow{1}{9mm}[20mm]{\small $ \frac{\Delta M}{M} [\mbox{\%}]$} & \includegraphics[width=4.0 cm , height=3.4 cm]{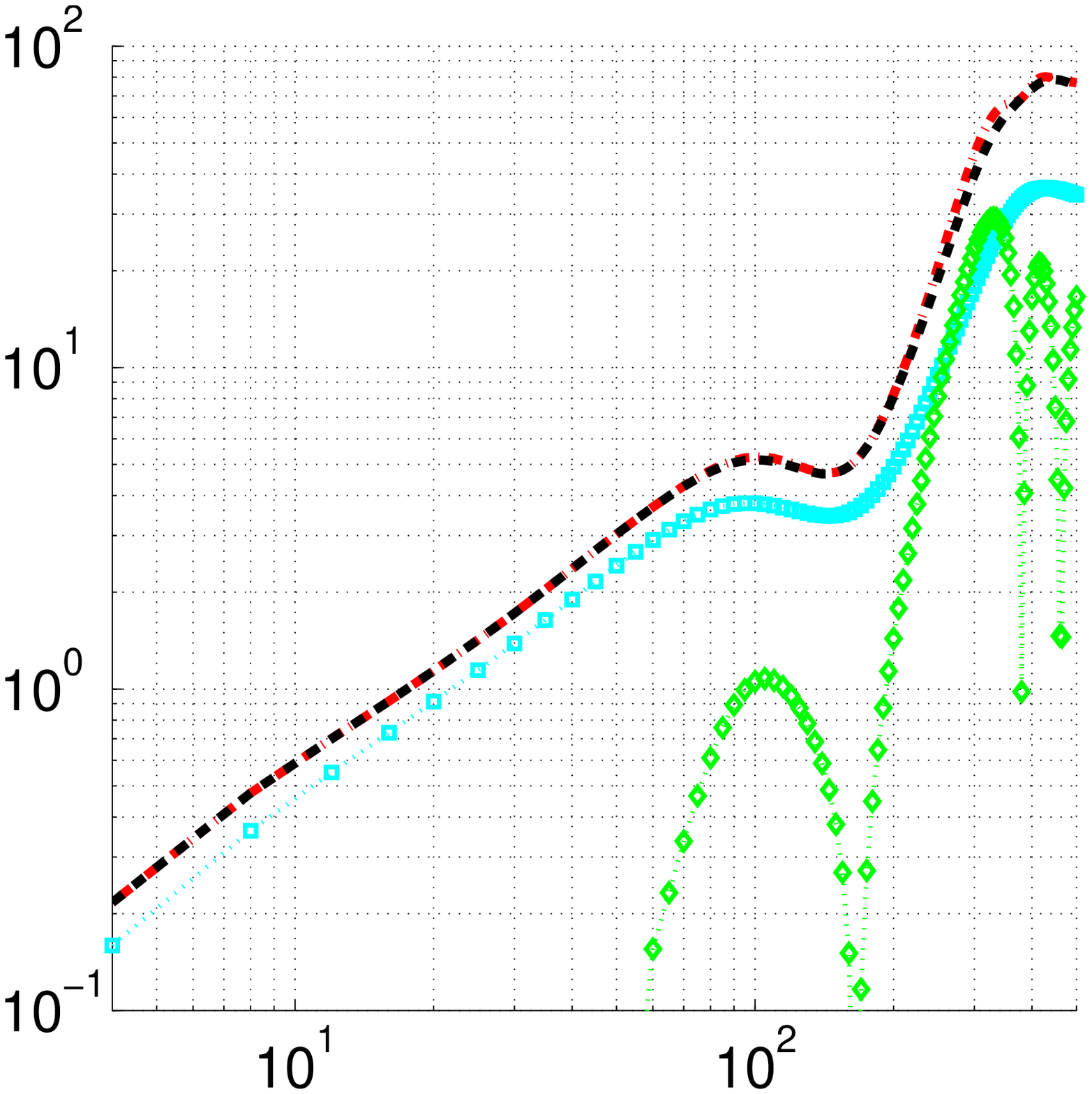} & \hskip -.2cm \multirow{1}{7.5mm}[20mm]{\small $ \frac{\Delta \eta}{\eta} [\mbox{\%}]$} & \includegraphics[width=4.0 cm,  height=3.4 cm]{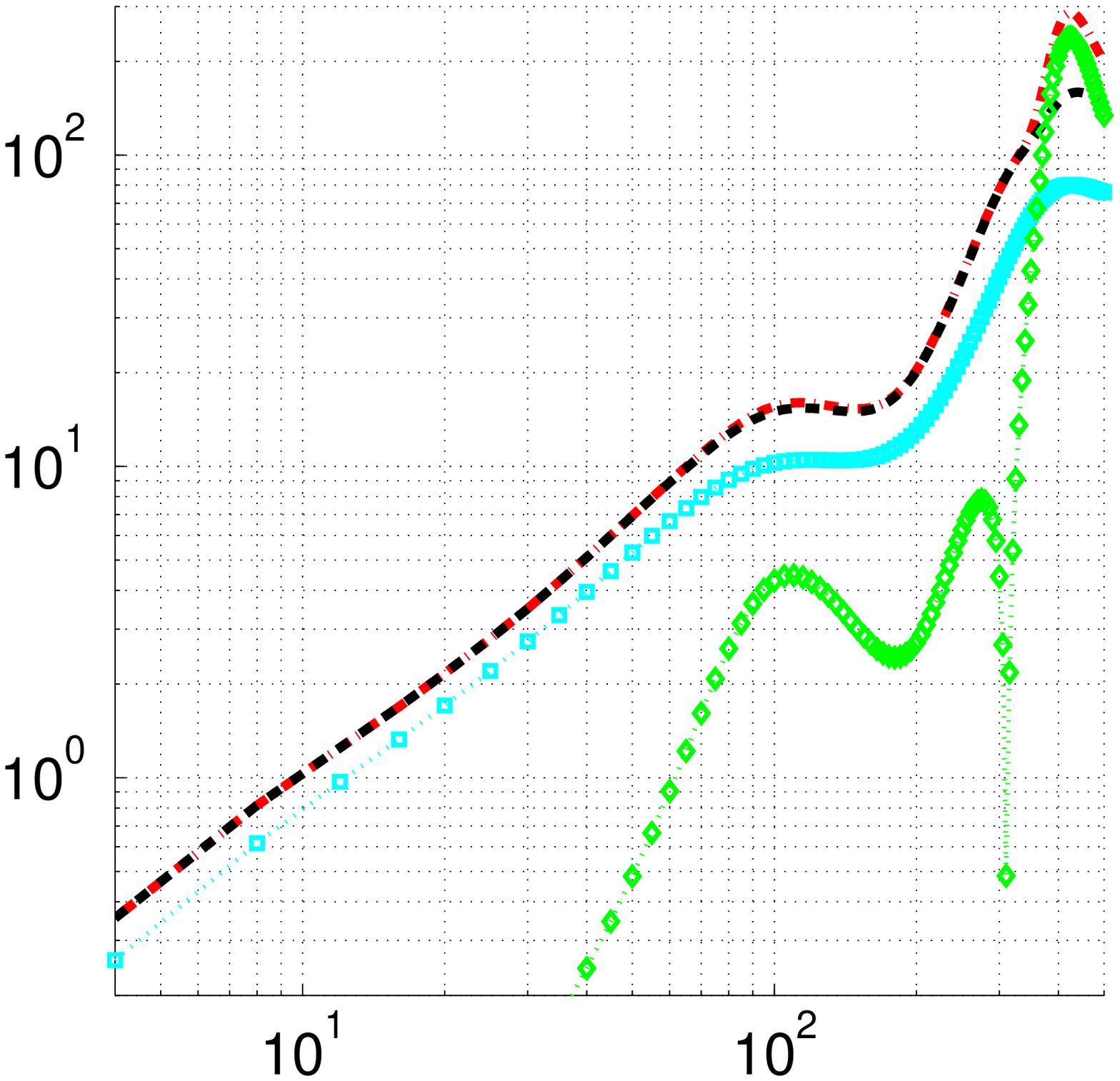} \\
\hskip -.2cm \multirow{1}{6mm}[20mm]{\small $ \Delta t [\mbox{s}]$} & \includegraphics[width=4.0 cm , height=3.4 cm]{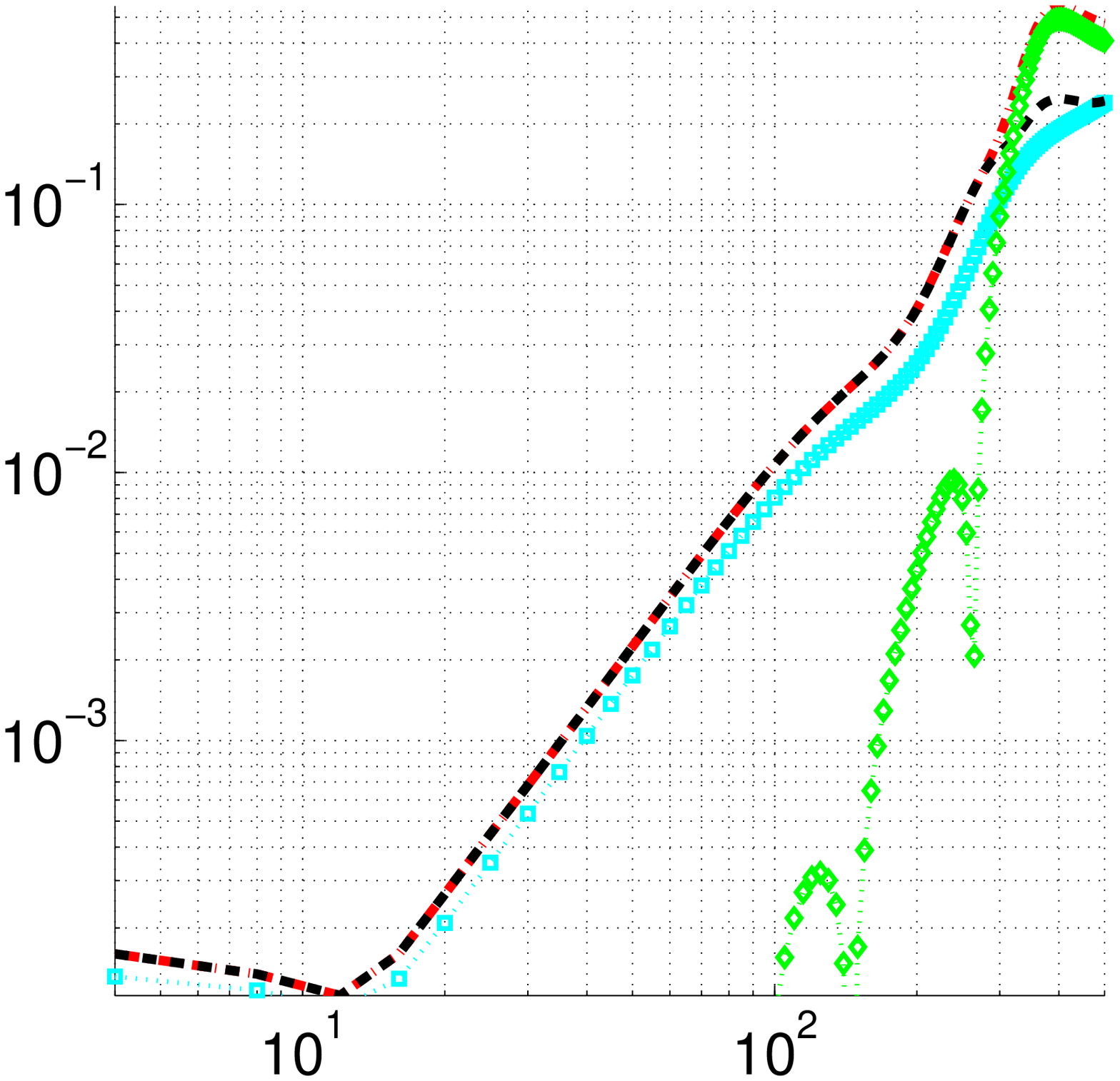} & \hskip -.2cm \multirow{1}{9mm}[20mm]{\small $ \frac{\Delta M}{M} [\mbox{\%}]$} & \includegraphics[width=4.0 cm , height=3.4 cm]{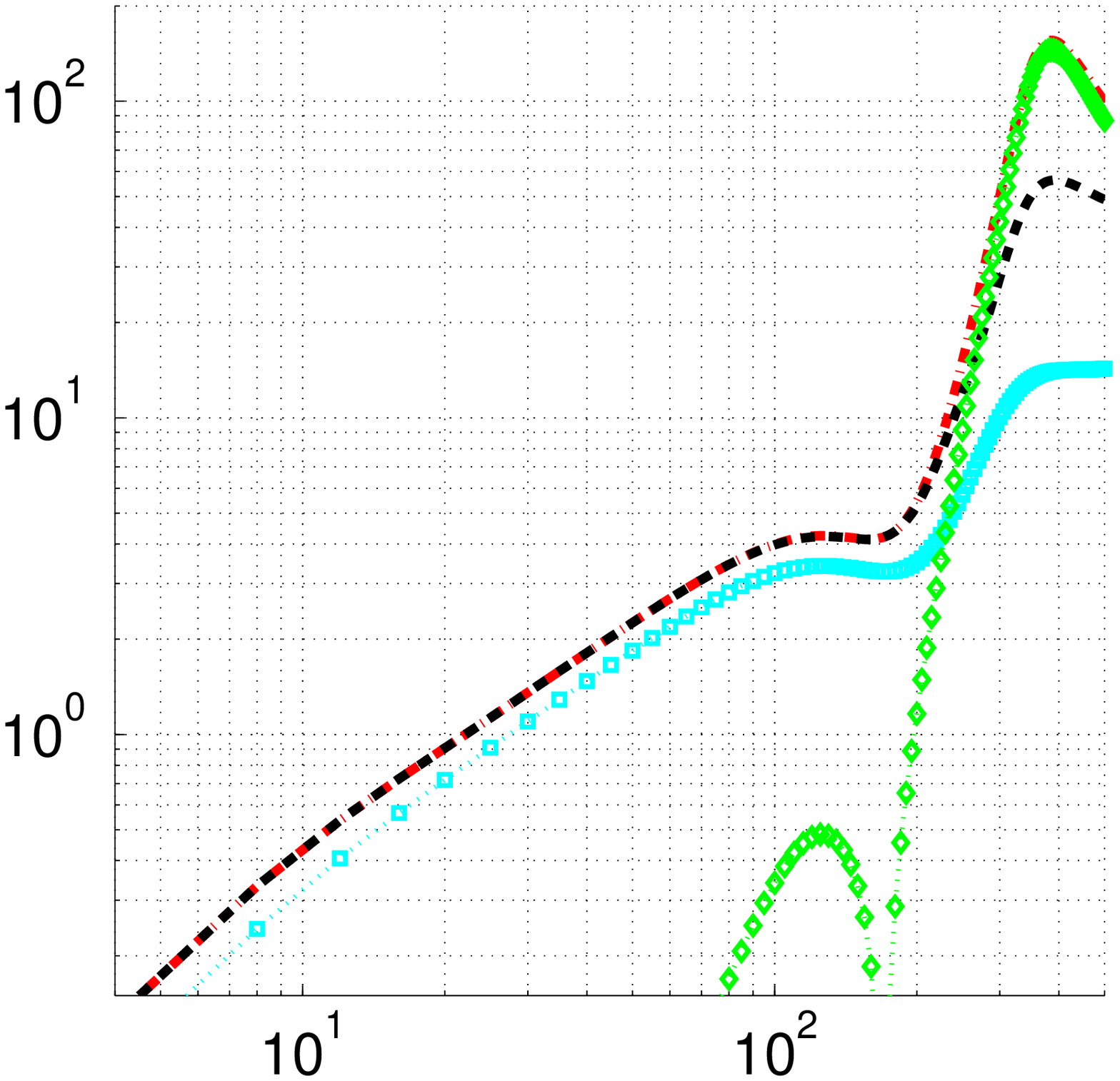} & \hskip -.2cm \multirow{1}{7.5mm}[20mm]{\small $ \frac{\Delta \eta}{\eta} [\mbox{\%}]$} & \includegraphics[width=4.0 cm,  height=3.4 cm]{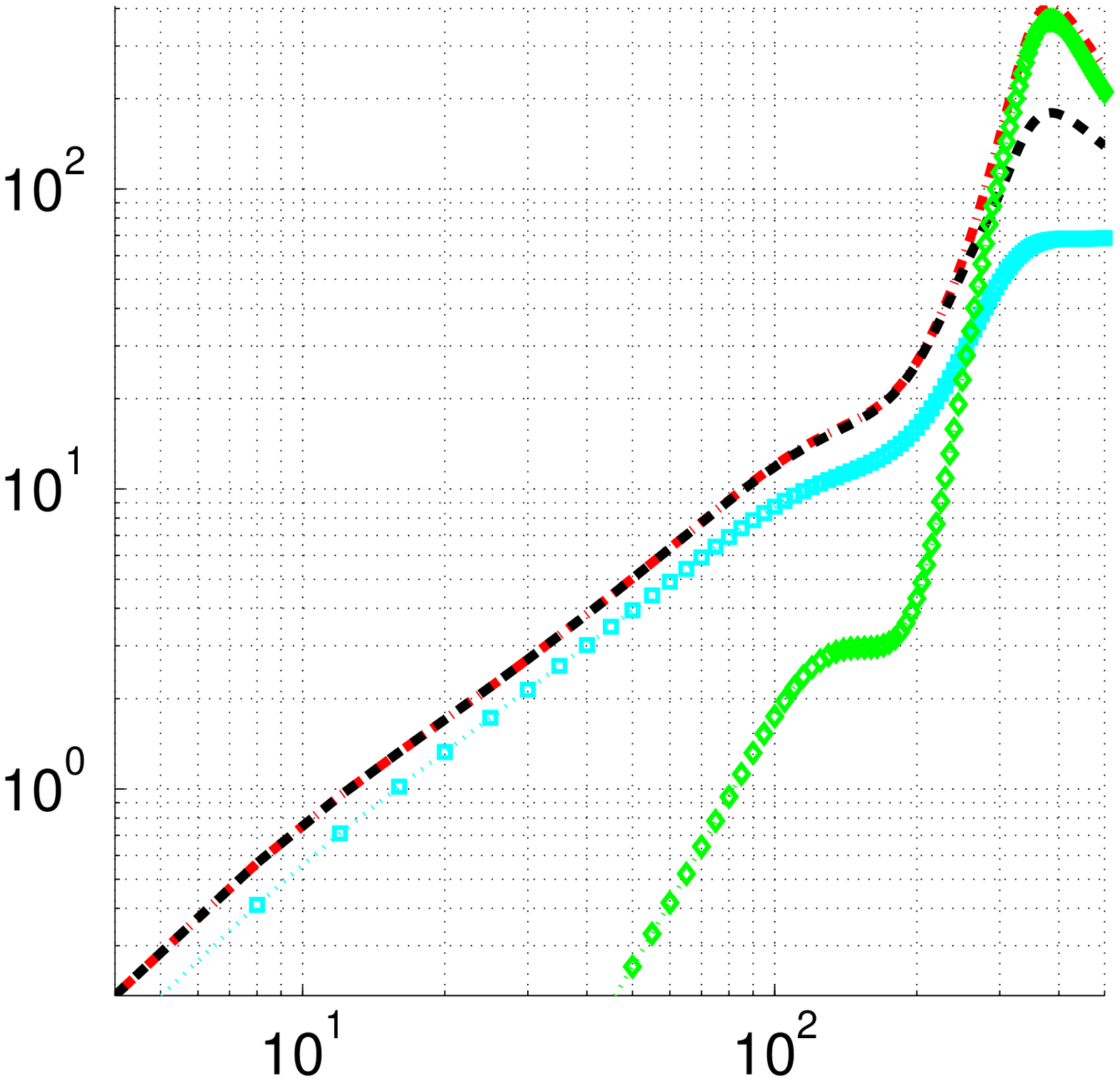} \\
&\hskip 10mm{ {\small $M$}}\quad & &\hskip 10mm{ {\small $M$}}\quad & &\hskip 10mm{ {\small $M$}}\quad 
\end{tabular}
\caption{\small (Color Online) The same as fig. \ref{AgainstMAL}, except that the Advanced Virgo noise is used.}\label{AgainstMAV}
\end{figure}
\end{widetext}
% \clearpage
% \vskip 6 cm
% $$$$
% \clearpage
% 
% \begin{figure}
% \centering
%  \begin{tabular}{ll}
% \multirow{1}{15mm}[20mm]{\small $ bias_{\frac{\Delta M}{M}} [\mbox{\%}]$} & \includegraphics[width=4.3 cm]{./plots/BIAS-ONLY-dM_eta22AL.eps} \\
% \quad \qquad & \qquad\qquad\quad { {\small    $M$}}\quad 
% \end{tabular}\caption{{\footnotesize	 (Color Online) The bias, first order (red dashed), second order (cyan dotted) and total (blue line) using the AdvLigo noise, for a fixed value of $\rho=10$, and with $\eta=0.2222$}\label{Plotminima}}
% \end{figure}
%\caption{ AdvVirgo}

\section{Conclusions and future work}
In this work we apply a recently derived methodology to the errors in 
estimating physical parameters from IMR signals. The asymptotic
expansion of the bias and the covariance are critical to have 
realistic estimates of the error for SNR below 20 
where the first detections of present and future laser interferometers 
might live. The behaviour of the errors, in terms of minima and 
maxima, in different regions of the parameter space appear to be more 
elaborate than predicted by the CRLB. For example the bias can become 
dominant for MLEs on sytems with large masses. This paper 
will aid the preparatory work that the scientific community 
is undertaking to prepare for the scientific runs of the 
advanced version of the earth based laser interferometers.

\begin{figure}[htb]
\begin{tabular}{ll}
\multirow{1}{15mm}[20mm]{\small $ bias_{\frac{\Delta M}{M}} [\mbox{\%}]$} & \includegraphics[width=4.3 cm]{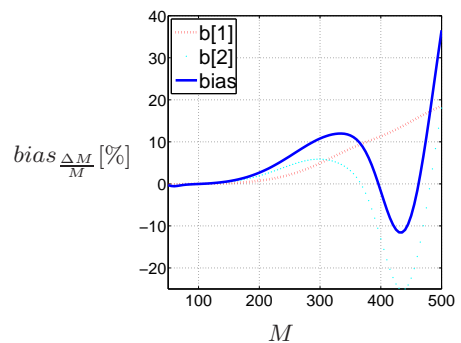} \\
\quad \qquad & \qquad\qquad\quad { {\small    $M$}}\quad 
\end{tabular}\caption{{\footnotesize (Color Online) The bias, first order (red dashed), second order (cyan dotted) and total (blue line) using the AdvLigo noise, for a fixed value of $\rho=10$, and with $\eta=0.2222$}\label{PlotBias}}
\end{figure}
\clearpage
\section{Acknowledgments}

The authours would like to thank Alessandra Buonanno for discussions on th EOB waveforms. S.V. thanks Walter Del Pozzo and Chris Van Den Broeck for useful discussions. M.Z. thanks the National Science Fondation for the support through the awards NSF855567 and NSF0919034.


\begin{thebibliography}{99}
\bibitem{RATE} LIGO Scientific Collaboration and Virgo Collaboration, {\it ArXiv},  gr-qc:1003.2480 
\bibitem{adl} http://www.ligo.caltech.edu/advLIGO/
\bibitem{adl2} http://wwwcascina.virgo.infn.it/advirgo/
               \bibitem{bhev1} R. Narayan, {\it New Journal of Physics},{\bf  7}, 199 (2005).
\bibitem{bhev2} J.Kormendy and D.Richstone,{\it Annu. Rev. Astron. Astrophys}, {\bf 33}, 581 (1995).
\bibitem{bhev3} R. Schodel et al., {\it Nature},{\bf 419}, 694 (2002).
\bibitem{bhev4} M.C. Miller and E.J.M. Colbert, {\it Int J. Mod. Phys. D}, {\bf 13}, 1 (2004).
\bibitem{aji5}S.Komossa et al., {\it Astrophys. J.}, {\bf 582}, L15 (2003)
\bibitem{aji6}L.Ballo et al, {\it Astrophys. J.}, {\bf 600}, 634 (2004)
\bibitem{aji7}M Guainazzi, E. Piconcelli, E. Jimenez-Bailon, and G. Matt, {\it Astron. Astrophys.},{\bf 429}, L9 (2005)
\bibitem{aji8}D.A.Evans et al., {\it Arxiv}, astro-ph/0712.2669 (2007)
\bibitem{aji9}S.Bianchi, M. Chiaberge, E. Piconcelli, M. Guainazzi and G.Matt, {\it Arxiv}, astro-ph/0802.0825 (2008)
\bibitem{aji10}K.A. Postnov and L.R. Yungelson, {\it Living Rev.Relativity},{\bf 9}, (2006)
\bibitem{aji11} P. Amaro-Seoane and M.Freitag, {\it Astrophys. J.},{\bf 653}, L53 , astro-ph/0610478, (2006)
\bibitem{aji12}J.M. Fregeau, S.L. Larson, M.C. Miller, R.O'Shaughnessy, and F.A.Rasio,{\it  Astrophys. J}, {\bf 646} ,L135 (2006)
\bibitem{aji13}I. Mandel, D.A. Brown, J.R. Gair, and M.C. Miller, {\it ArXiv}, astro-ph:0705.0285
\bibitem{Buonanno2009} A. Buonanno, Y. Pan, H.P. Pfeiffer, M.A. Scheel, L. Buchman and L.E. Kidder, {\it ArXiv}, gr-qc/0902.0790 (2009)
\bibitem{Buonanno2007} A. Buonanno, Y. Pan, J.G. Baker, J. Centrella, B.J. Kelly, S.T. McWilliams and R. Van Meter, {\it ArXiv} gr-qc/0706.3732 (2007)
\bibitem{Damour2009} T. Damour and A. Nagar, {\it ArXiv} gr-qc/0902.0136 (2009)
\bibitem{Santamaria2010} L. Santamaria, F. Ohme, P. Ajith, B. Bruegmann, N. Dorband, M. Hannam, S. Husa, P. Moesta, D. Pollney, C. Reisswig, E.L. Robinson, J. Seiler and B. Krishnan, {\it ArXiv} gr-qc/1005.3306 (2010)
\bibitem{ninja} The ninja collaboration: https://www.ninja-project.org/doku.php
\bibitem{IMR-ZanolinVitale2010} M. Zanolin, S. Vitale and N. Makris, {\it Phys. Rev .D}, {\bf 81} 124048  (2010)
\bibitem{IMR-Helstrom1968} C.W. Helstrom, {\it Statistical Theory of Signal Detection}, International Series of Monographs in Electronics and Instrumentation, {\bf vol. 9}, (Pergamon Press, Oxford; New York, 1968), 2nd edition.
\bibitem{IMR-Blanchet2006} L. Blanchet, {\it Living Rev. Relativity}, {\bf 9} (2006)
\bibitem{IMR-Pretorius2005} F. Pretorius, {\it Phys. Rev. Lett.} {\bf 95}, 121101 (2005)
\bibitem{IMR-Campanelli2006} M. Campanelli, C.O. Lousto, P. Marronetti, and Y. Zlochower, {\it Phys. Rev. Lett.} {\bf 96} 111101 (2006).
\bibitem{IMR-Baker2006} J.G. Baker, J. Centrella, D.I. Choi. M. Koppitz 	and J. van Meter, {\it Phys. Rev. Lett.}{\bf 96} 111102 (2006)
\bibitem{IMR-Herrmann2007} F. Herrmann, I. Hindler, D. Shoemaker and P. Laguna, {\it Class. Quantum Grav.}, {\bf 24}, S33 (2007)
\bibitem{IMR-Sperhake2006} U. Sperhake, {\it ArXiv}: gr-qc/0606079
\bibitem{IMR-Brugmann2006} B. Br\"{u}gmann, J.A. Gonz\'{a}lez, M. Hannam, S. Husa, U Sperhake and W. Tichy, {\it ArXiv}:gr-qc/0610128 (2006)
\bibitem{IMR-Thornburg2007} T. Thornburg, P. Diener, D. Pollney, L. Rezzola, E. Schnetter, E. Seidel and R. Takahashi, {\it Class. Quantum Grav. }{\bf 24} 3911 (2007)
\bibitem{IMR-Etienne2007} Z.B. Etienne, J.A. Faber, Y.T. Liu, S.L. Shapiro and T.W. Baumgarte, {\it ArXiv}:gr-qc/0707.2083 (2007)
\bibitem{IMR-Schutz2009} B.S. Sathyaprakash and B.F. Schutz, {\it Living Rev.Relativity}, {\bf 12} (2009)
\bibitem{IMR-Vecchio1998} D. Nicholson, A. Vecchio, Phys. Rev. D \textbf{57}, 4588 (1998)
\bibitem{IMR-Bala1996} R. Balasubramanian, B.S. Sathyaprakash and S.V. Dhurandhar, Phys. Rev. D {\bf{53}}, 3033 (1996)
\bibitem{IMR-Bala1998} R. Balasubramanian and S.V. Dhurandhar, {\it Phys. Rev. D}, {\bf 57}, 3408 (1998)
\bibitem{IMR-Cokelaer2008} T. Cokelaer, {\it Class. Quantum Grav.} {\bf 25}, 184007 (2008)
\bibitem{IMR-Ajith2007} P. Ajith et al., {\it Class. Quantum Grav.}, {\bf 24}, S689 (2007)
\bibitem{IMR-Ajith2008} P. Ajith,  {\it Class. Quantum Grav.}, {\bf 25},114033 (2008)
\bibitem{IMR-AjithApril2009} P. Ajith and S. Bose, {\it arXiv}, gr-qc/0901.4936 (2009)
\bibitem{IMR-AjithMay2009} P. Ajith et al., {\it ArXiv}, gr-qc/0710.2335
\bibitem{IMR-AjithSept2009} P. Ajith, M. Hannam, S. Husa, Y. Chen, B. Br\"{u}gmann, N. Dorband, D. M\"{u}ller, F. Ohme, D. Pollney, C. Reisswig, L. Santamaria and J Seiler, {\it arXiv}, gr-qc/0909.2867v1 (2009)
\bibitem{IMR-Cutler1994} C. Cutler and E.E. Flanagan, {\it Phys. Rev. D}, {\bf 49}, 6, 2658 (1994)
\bibitem{IMR-Arun2005} K.G Arun, B.R. Iyer, B.S. Sathyaprakash and P.A. Sundararajan, {\it arXiv}, gr-qc/0411146 (2005)
\bibitem{Abadie2010} J. Abadie et al, {\it Class. Quantum Grav.}, {\bf 27}, 173001 (2010)
\end{thebibliography}
 \end{document}